\documentclass[lettersize,journal]{IEEEtran}
\usepackage{amsmath,amsfonts}
\usepackage{algorithmic}
\usepackage{array}
\usepackage{textcomp}
\usepackage{stfloats}
\usepackage{url}
\usepackage{verbatim}
\usepackage{graphicx}
\usepackage{makecell}
\usepackage{rotating}
\usepackage[tight]{subfigure}

\usepackage{booktabs}       
\usepackage{xcolor}         
\usepackage{multirow}

\def\BibTeX{{\rm B\kern-.05em{\sc i\kern-.025em b}\kern-.08em
    T\kern-.1667em\lower.7ex\hbox{E}\kern-.125emX}}
\usepackage{balance}
\begin{document}
\title{Smaller Is Bigger: Rethinking the Embedding Rate of Deep Hiding}
\author{Han Li, \and Hangcheng Liu, \and Shangwei Guo, \and Mingliang Zhou, \and Ning Wang, \and Tao Xiang, and Tianwei Zhang
\thanks{Han Li and Hangcheng Liu contribute equally.}
\thanks{Han Li, Hangcheng Liu, Shangwei Guo, Mingliang Zhou, Ning Wang, and Tao Xiang are with College of Computer Science, Chongqing University, China.}
  \thanks{Tianwei Zhang is with School of Computer Science and Engineering, Nanyang Technological University, Singapore.}
}



\maketitle


\begin{abstract}
  Deep hiding, concealing secret information using Deep Neural Networks (DNNs), can significantly increase the embedding rate and improve the efficiency of secret sharing. Existing works mainly force on designing DNNs with higher embedding rates or fancy functionalities. In this paper, we want to answer some fundamental questions: how to increase and what determines the embedding rate of deep hiding. To this end, we first propose a novel Local Deep Hiding (LDH) scheme that significantly increases the embedding rate by hiding large secret images into small local regions of cover images. Our scheme consists of three DNNs: hiding, locating, and revealing. We use the hiding network to convert a secret image in a small imperceptible compact secret code that is embedded into a random local region of a cover image. The locating network assists the revealing process by identifying the position of secret codes in the stego image, while the revealing network recovers all full-size secret images from these identified local regions. Our LDH achieves an extremely high embedding rate, i.e., $16\times24$ bpp and exhibits superior robustness to common image distortions. We also conduct comprehensive experiments to evaluate our scheme under various system settings. We further quantitatively analyze the trade-off between the embedding rate and image quality with different image restoration algorithms.
\end{abstract}


\section{Introduction}\label{sec:introduction}
\IEEEPARstart{D}{ata} hiding or steganography is the art of hiding secret messages into innocent-looking cover objects such as images, videos or other computer files \cite{morkel2005overview,cheddad2010digital,anderson1998limits,bandyopadhyay2008tutorial,liu2021new,guan2022double}. This technology allows secret sharing in public channels without drawing suspicion to the transmission of the hidden information. A data hiding scheme usually consists of two phases, hiding and revealing. In the hiding phase, the secret message owner embeds the message into a cover medium (e.g., an image) to produce a stego medium, while preserving the fidelity of the cover medium. In the revealing phase, the receiver extracts the secret embedded message from the stego medium. In this paper, we concentrate on hiding and revealing secret message in images.

During the evaluation of data hiding schemes, the embedding rate is one of the most important metrics to measure the efficiency of secret sharing, which can be quantified by bits per pixel (bpp). Many efforts have been made to improve the embedding rate. For example, HUGO \cite{pevny2010using} minimizes a suitably-defined distortion with a designed coding algorithm to increase the embedding rate. Unfortunately, such traditional data hiding schemes \cite{pevny2010using, chan2004hiding,tamimi2013hiding,li2007steganographic,raja2005secure,mckeon2007strange,denemark2015improving,li2014new,li2015strategy} usually have a low embedding rate, e.g., HUGO only achieves about 0.5 bpp. Such a low embedding rate indicates that HUGO can be only used to transfer short text messages. To the best of our knowledge, this is also a common problem for most traditional data hiding schemes \cite{chan2004hiding,tamimi2013hiding,denemark2015improving,li2014new,li2015strategy}, especially for frequency-domain based schemes \cite{li2007steganographic,raja2005secure,mckeon2007strange}.


Recently, deep learning techniques have been applied to data hiding (i.e., deep hiding), which hides and reveals secret messages using deep neural networks (DNNs) and can significantly improve the embedding rate. Baluja et al. \cite{baluja2017hiding} first leverage DNNs to achieve a high embedding rate, 24 bpp, by hiding a full-size image into another one. After that, more deep hiding schemes \cite{zhang2020udh,yu2020attention,zhang2019invisible,lu2021large} are proposed to reach a similar or higher embedding rate. For example, Zhang et al. \cite{zhang2020udh} proposed a deep hiding scheme that can hide three secret images into a cover image (i.e., $3\times 24$ bpp) by training three pairs of hiding and revealing networks.


Existing deep hiding schemes have achieved a high embedding rate, which either improve the image quality of secret images \cite{yu2020attention,zhang2019invisible} or provide fancy functionalities such as hiding hyperlinks into an image \cite{tancik2020stegastamp}, light field messaging \cite{wengrowski2019light}. However, there is little analysis of why this happens and what affects the embedding rate. In this paper, we intend to answer two more fundamental questions: 1) \textit{why DNNs can greatly improve the embedding rate?} 2) \textit{can the embedding rate be further greatly improved?} and 3) \textit{Are there any trade-offs between the embedding rate and image quality?}.


For the first question, we find that DNNs are trained to encode secret images into imperceptible codes that are further embedded into cover images. Such an efficient coding technique can effectively reduce the scale of the secret images and highly improve the embedding rates of the corresponding hiding schemes. With such insights, we address the second question by designing a novel Local Deep Hiding (LDH) and further increase the embedding rate to an amazing level, $16\times 24$ bpp. In particular, our LDH uses a carefully designed DNN to encode a full-size secret image into a small \textit{imperceptible compact secret code} that is the key to improving the embedment rate. The small imperceptible compact secret code is randomly embedded in to a local region of a cover image. Since the remaining pixels of the cover image can be used for hiding other secret images, our LDH can embed much more images than existing deep hiding schemes that embed a full-size secret image into all pixels of the cover image.

Our LDH consists of three DNNs: hiding, locating, and revealing networks. The hiding network converts a given secret image into a small imperceptible compact secret code (looks like a noise image of small size) that will be added to a random local region of a cover image to produce the corresponding stego image. The locating network identifies the positions of all embedded secret codes in the stego image and outputs a binary location map, with which the revealing network recovers all full-size secret images from the determined small local regions in the stego image. Our LDH also can complete the embedding and recovering for all secret images through the same hiding and revealing networks, instead of training multiple pairs of hiding and revealing networks. Experimental results show that our LDH can achieve a much higher embedding rate while preserving the quality of both the cover and secure images.

To answer the last question, we construct an experiment scenario of secret sharing and present an in-depth analysis from the perspective of the message receiver. In particular, we assume that the receiver can use different algorithms to enhance the quality of the revealed secret images. We measure the highest embedding rates that can be achieved by various deep hiding schemes when the quality reaches a certain threshold. We show that the embedding rates of deep hiding schemes varies with different quality threshold. And as the embedding rate increases, the quality of the results revealed by existing hiding schemes gradually decreases, while our LDH does not exhibit this phenomenon. Besides, the embedding rate of a deep hiding scheme may be increased if the receiver owns a powerful image restoration model to enhance the image quality of the secret images.

We summarize our main contributions below:
\begin{itemize}
    \item We propose a novel Local Deep Hiding (LDH) scheme to encode secret images into small imperceptible compact secret codes and further significantly increase the embedding rate.
    \item We propose a two-stage training strategy to enhance the performance of our LDH.
    \item We conduct extensive experiments to evaluate the superiority of our LDH.
    \item We present an in-depth study to quantitatively analyze the trade-off between the embedding rate and image quality.
\end{itemize}

The remainder of the paper is organized as follows: Section \ref{sec:related_work} introduce data hiding and related works. In Section \ref{sec:algorithm}, we specify the details of our proposed scheme. In Section \ref{sec:experimental_evaluation} evaluates our scheme under various system settings, followed by a quantitative analysis of the trade-off between the embedding rate and image quality. Section \ref{sec:conclusion} concludes this paper.
\section{Background and Related Work}
\label{sec:related_work}
To facilitate the description of a data hiding scheme, we use $S$ and $C$ to represent a secret message and a cover image respectively. $\epsilon_1$ and $\epsilon_2$ are two hyperparameters. $D$ measures the difference between two given inputs. A data hiding scheme consists of two necessary algorithms:
    \begin{itemize}
        \item \textbf{Hiding algorithm} ($H$) embeds $S$ into $C$ without affecting the visual quality of $C$, and outputs a stego image $C'$ (i.e., $C'=H(C, S)$), which has $D(C',C)\leq\epsilon_1$;
        \item \textbf{Revealing algorithm} ($R$) reveals the embedded secret message from $C'$ as $S'=R(C')$ and needs to meet $D(S',S)\leq\epsilon_2$.
    \end{itemize}
The performance of a data hiding scheme is often measured by the number of bits a cover pixel can carry, i.e., bits per pixel (bpp). A higher numerical value of bpp means that the corresponding data hiding scheme has a higher message capacity (i.e., a higher embedding rate).

\subsection{Data Hiding of Low Embedding Rates}
Some data hiding schemes embed secret messages into frequency domain domains such as iscrete cosine transform (DCT) domain \cite{provos2003hide,fard2006new,attaby2018data}, discrete wavelet transform (DWT) domain \cite{reddy2009high,atawneh2017secure}, discrete Fourier transform (DFT) domain \cite{mckeon2007strange,ruanaidh1996phase} and integer wavelet transform (IWT) domain \cite{gulve2015image,kalita2019new}. But such frequency-domain data hiding schemes can only reach a low embedding rate, such as 2 bpp or lower.
For example, Attaby et al. \cite{attaby2018data} proposed DCT-M3 algorithm, which uses modulus 3 of the difference between two DCT coefficients to embed two bits of the secret message. Atawneh et al. \cite{atawneh2017secure} applied a diamond encoding method to realize a DWT-based hiding scheme, enhancing the quality of the stego images. McKeon and Robert \cite{mckeon2007strange} proposed a DFT-based scheme that provides a sophisticated method for hiding data in movies. Kalita et al. \cite{kalita2019new} developed an IWT-based method that uses only high-frequency subbands to conceal secret messages for resisting steganalysis. Although frequency-based schemes often show good robustness against common interference operations (e.g., compression), but they are highly computational complexity and can only achieve limit embedding rates.

Instead, other data hiding schemes embed secret messaves into the spatial domain, such as
least significant bit (LSB) \cite{chan2004hiding,jung2015steganographic,sahu2020reversible}, pixel value differencing (PVD) \cite{mandal2022high,li2018steganography}, histogram shifting \cite{thodi2007expansion,tai2009reversible,xie2020adaptive}, expansion based \cite{tian2003reversible,gujjunoori2019difference}, multiple bit-planes \cite{kieu2009high,nyeem2017reversible}. These methods have higher embedding rates than frequency-domain based methods, but are usually less robust and less resistant to steganalysis \cite{fridrich2001reliable,dumitrescu2002detection,fridrich2004estimation,ker2004improved,ker2005general}. The spatial-domain methods perform pixel operations directly and thus usually have higher embedding rates (such as more than 2 bpp). Recent works also use deep learning techniques to improve the embedding rate. For example,
Hayes et al. \cite{hayes2017generating} devised a GAN-based data hiding to enhance the quality of the stego image. Ghamizi et al. \cite{ghamizi2019adversarial} used deep neural networks and adversarial attacks to embed secret information into images and improves the embedding rate to some extent. Hu et al. \cite{hu2018novel} proposed a steganography without embedding (SWE) method, which uses a trained generator neural network model to generate the stego image based on a noise vector, to further improve the ability to resist detection algorithms. Ray et al. \cite{ray2021image} used a deep supervision based edge detector to embed more bits into the edge pixels. Tancik et al. \cite{tancik2020stegastamp} devised a new framework StegaStamp to realize message hiding and revealing in the physical world. However, these data hiding schemes with deep learning focused on embedding binary messages into cover images and also have low embedding rates.


\subsection{Data Hiding of High Embedding Rates}
A higher embedding rate indicate that more secret messages can be delivered through the same cover image, increasing the effectiveness of secret sharing. Existing data hiding schemes of high embedding rates hide full-size images within images using DNNs, which can realize more than 24 bpp.
According to the different model structures, we can classify these schemes into three categories: encoder-decoder structure \cite{baluja2017hiding,weng2019high,rahim2018end,zhang2019invisible,zhang2020udh,das2021multi}, generative adversarial network (GAN) structure \cite{zhang2019steganogan,meng2019steganography,yu2020attention}, and invertible neural network structure (INN) \cite{lu2021large,guan2022deepmih}.

\textbf{Encoder-Decoder Structure.} Since the data hiding and revealing processes correspond exactly to the encoding and decoding processes, encoder-decoder structure is suited for data hiding models.
For example, Baluja et al. \cite{baluja2017hiding} first designed a framework for hiding images within image, which uses a preprocessing network to embed the secret images into a cover image using a hiding network.
Zhang et al. \cite{zhang2020udh} removed the preprocessing network and designed a framework for large-capacity data hiding, in which the encoding process of secret images is independent of the choice of cover images and the secret codes can be hidden in any cover image. They further trained three pairs of hiding and revealing networks to improve the embedding rate to 72 bpp.
Similarly, Das et al. \cite{das2021multi} proposed MISDNN, which adopts several pairs of preparing networks, hiding networks, and revealing networks to achieve multiple-image hiding.

\textbf{GAN Structure-Based.} In this type of deep hiding, both the hiding and revealing processes can be achieved by a specialized generator, while the discriminator is responsible for making the generated results as close as possible to the original ones.
Meng et al. \cite{meng2019steganography} adopted CycleGAN \cite{isola2017image} to achieve hiding images within images and improved the security of data hiding by adversarial training. Similarly, Yu et al. \cite{yu2020attention} also adopted GAN in deep hiding and further introduced an attention mechanism to improve the image quality.

\textbf{INN Structure.} This type of deep hiding schemes utilizes INNs \cite{dinh2014nice,dinh2016density} and shares parameters in the hiding and revealing networks instead of training two independent networks.
Lu et al. \cite{lu2021large} used a INN to play both the hiding and revealing networks simultaneously and realized multiple-image hiding by concatenating multiple secret images in the channel dimension.
Guan et al. \cite{guan2022deepmih} developed an invertible hiding neural network (IHNN) that can be cascaded multiple times. When hiding multiple secret images, IHNN is able to guide the current image hiding based on the previous image hiding results.

Although existing deep hiding schemes improve the hiding performance significantly, especially the embedding rate, they still cannot fully answer the fundamental questions. To further analyze the functionality and improve the embedding rate, we will propose a novel local deep hiding scheme below.

\section{Methodology}\label{sec:algorithm}

\textbf{Insight.}
While surprised by the high embedding rate of deep hiding, we are also thinking about the reasons behind such improvement in the embedding rate and how to further improve it. By comparing traditional data hiding with deep hiding, we find that the encoding ability of DNNs is critical to improve the embedding rate since images are of high redundancy and DNNs can automatically embed secret image's information into the cover image without affecting its fidelity.

To demonstrate this phenomenon, we design a local embedding strategy to explain the working principle of deep hiding, which can also achieve a much higher embedding rate without affecting the visual quality of cover images. In particular, we propose a novel local deep hiding (LDH) method, where we convert a secret image into a small imperceptible compact code (e.g., the size of the code can be only $\frac{1}{16}$ of the secret image's size) using our hiding network. And we require the revealing network to recover the corresponding full-size and complete secret image. Obviously, the small imperceptible compact code carries the full information of the secret image and can be recognized by the revealing network, which is difficult for traditional data hiding due to their limited encoding ability. In addition, with the secret code becoming smaller, we can hide more images within different local regions of a cover image to improve the embedding rate by carefully designing the losses and the training process.

\begin{figure*}[t]
    \centering
    \includegraphics[width=1.0\linewidth]{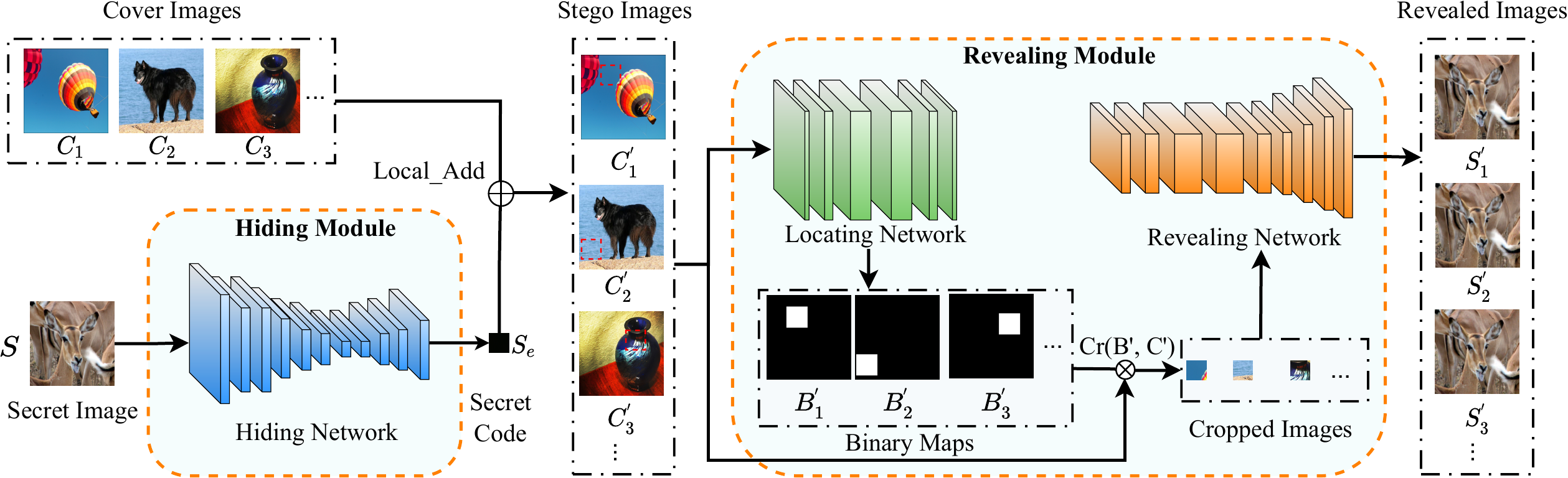}
    \caption{The proposed LDH framework.}
    \label{fig:pipeline}
\end{figure*}

\textbf{Pipeline.} We illustrate the pipeline of the proposed LDH in Fig. \ref{fig:pipeline}, which consists of two modules: hiding and revealing module. \textit{In the hiding module,} we design a hiding network ($H$) to convert a given secret image ($S$) into a small imperceptible compact code ($S_e$), whose size is smaller than $S$. And then, we hide the compact secret code $S_e$ within a random local region of $C$ to produce the corresponding stego image ($C'$). Note that our method is  universal, where the same $S_e$ can be hidden into different cover images. \textit{In the revealing module,} there are two DNNs: locating and revealing networks. We first design a locating network $P$ to identify the position of $S_e$ in $C'$ by producing a binary map ($B'$). Next, we crop the determined local region of $C'$ according to $B'$ and feed the cropped result to a revealing network ($R$) to recover the corresponding full-size secret images ($S'$).

\subsection{Networks for LDH}
\textbf{Hiding Network.} Existing deep hiding schemes \cite{baluja2017hiding,zhang2020udh,yu2020attention} usually conceal one or more secret image(s) within a cover image of the same size. However, since multiple secret image information is embedded into all pixels of the cover image, the robustness and the visual quality of the revealed secret images would be affected. Thus, it is difficult to increase the embedding rate of deep hiding in such a global manner. In addition, one cannot clearly figure out the working principle of deep hiding. To address the problems, we consider local embedding and design a novel hiding network to covert full-size secret images into small imperceptible compact codes as we have illustrated in Fig. \ref{fig:pipeline}.

In the hiding network, we first adopt some down-sampling layers to reduce the scale of $S$, which are used to produce intermediate representations of $S$. And then, these intermediate representations are further encoded by $H$ that outputs the final small imperceptible compact secret code $S_e$. We use a scaling factor $\omega$ to control the size of $S_e$ in this paper, where $\omega$ is the ratio of the width of $S_e$ to the width of $S$. After obtaining $S_e$, we hide it within a random local region of a given cover image $C$ through element-wise addition as shown in Fig. \ref{fig:pipeline}. The complete hiding process can be formally represented as
\begin{equation}
    C'=Local\_Add(C,H(S, \omega)),
    \label{eq:local_add}
\end{equation}
where $Local\_Add$ denotes the addition operation. Please note that $\omega$ determines the upper bound of embedding rate.

\textbf{Locating Network.} Since the small imperceptible code $S_e$ is embedded into $C$ locally and randomly, we need to identify the position of $S_e$ in $C'$. To this end, we design a locating network $P$ to automatically identify the regions that are embedded with $S_e$, which will assist the revealing network to reconstruct the secret image. The locating network $P$ takes $C'$ as its input and determines the position of $S_e$ in $C'$ by generating a binary map $B'$. In particular, we use 1 in $B'$ to indicate the corresponding pixel of $C'$ belongs to an embedding region and use 0 to indicate the opposite. The whole locating process is
\begin{equation}
    B'=P(C').
\end{equation}




\textbf{Revealing Network.} In previous works \cite{baluja2017hiding,zhang2019invisible,zhang2020udh,yu2020attention}, the revealing network recovers $S'$ from $S_e$ that has the same size as $S'$. Although these works achieve an embedding rate of 24 bpp or higher, they still do not fully exploit the potential of the revealing network. In LDH, we make full use of the capability of the revealing network to record extra image information besides $S_e$ in its parameters. So our revealing network can recover the full-size $S'$ from the small compact $S_e$. Specifically, in $R$, we use its bottom layers to pre-process $S_e$ for getting the corresponding intermediate representation while maintaining the size. Then, we use the top layers of $R$ to further convert the representation and increase its size gradually until meeting the original size of $S$.
The revealing process is summarized as
\begin{equation}
    S'=R(Cr(C',B')),
\end{equation}
where $Cr(C', B')$ means applying the cropping operation to $C'$ under the guidance of $B'$.

\subsection{Training Details}
We co-train the three networks to ensure the imperceptibility and security of secret sharing. In particular, we have three goals: 1) minimize the negative impact of the hiding to cover images; 2) accurately identify the location of the embedded regions; 3) minimize the difference between the original and revealed secret images. In the following, we introduce three loss functions to achieve these goals.

\textbf{Hiding Loss.} We use the hiding loss ($L_H$) for restraining $H$, which ensures the imperceptibility of $S_e$. A common way to make $C'$ close to $C$ is minimizing the $L_p$ distance between $C'$ and $C$ \cite{baluja2017hiding,zhang2020udh}. Therefore, in our experiments, we set $L_H$ as
\begin{equation}
    L_H = \|C-C'\|_p.
\end{equation}




\textbf{Locating Loss.} The accuracy of identifying the location of the embedded regions is crucial for our scheme since it affects the final revealing result directly. Before the training of $P$, we record the real positions of all embedded secret code $S_e$ in a binary map $B$ (i.e., ground truth), where $B_{ij}$ is 1 if the corresponding pixel in $C'$ carries secret information, otherwise it is 0. During the training process, we optimize $P$ to reduce the locating loss ($L_P$) between ground truth and the predict map $B'$ of $P$ , i.e.,
\begin{equation}
    L_P = \|B-B'\|_p.
\end{equation}


\textbf{Revealing Loss.} $R$ is responsible for revealing $S'$ from the cropped stego image $C'$, which should be as close as possible to the original secret image $S$. Therefore, we also design the corresponding revealing loss ($L_R$) in the form of $L_p$-norm as
\begin{equation}
    L_R = \|S-S'\|_p.
\end{equation}


With the above three loss items, we can obtain our final loss to training the three DNNs in our LDH, i.e.,
\begin{equation}
    L_T=\lambda_1 L_H + \lambda_2 L_P + \lambda_3 L_R,
\end{equation}
where $\lambda_1$, $\lambda_2$, $\lambda_3$ are three hyper-parameters.

\textbf{Training Strategy.} To better train the DNNs, we use a two-stage training strategy. Specifically, we split the whole training process into the pre-training and co-training phases. In the pre-training phase, we set $\lambda_2=0$. This is because, at the beginning of training, $P$ cannot determine the embedding regions correctly. We do not require $R$ to recover $S'$ from the non-embedding regions that would lead to a stagnation of training. Therefore, we do not consider $P$ in the pre-training phase. Note that, to train the $H$ and $R$, we use $B$ (the ground truth of locations) to crop $C'$ in this phase. After that, we start to co-train $H$, $P$, and $R$. Considering that $H$ and $R$ have been trained well, we focus primarily on training $P$ in the co-training phase. So we set $\lambda_2$ to a large value while $\lambda_1$ and $\lambda_2$ are small in the co-training.

\subsection{Multiple Image Hiding}
\begin{figure}[t]
    \centering
    \includegraphics[width=0.9\linewidth]{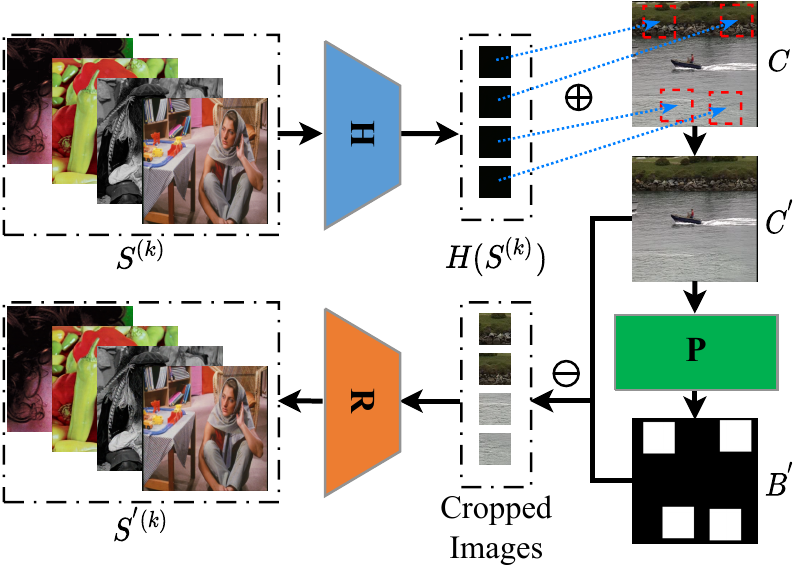}
    \caption{Concealing four different $S$ into the different positions of one $C$ using our LDH. All secret images could be detected and recovered in high quality.}
    \label{fig:diffpos}
    \vspace{-10pt}
\end{figure}

Benefiting from the local embedding strategy, LDH enables multiple image hiding for a higher embedding rate. As shown in Fig. \ref{fig:diffpos}, four secret images are encoded as the corresponding imperceptible compact codes by $H$ and then embedded into different local regions of $C$. Note that our LDH can also support hiding multiple images into the same or overlapped local regions, which only requires to considering and adding this type of hiding samples into the ground truth during the training process. Without loss of generality, we only consider embedding multiple images into different regions of the cover image. Our locating network $P$ can identify all the positions of the embedded secret images and outputs a binary map containing multiple white areas. Finally, $R$ recovers all full-size secret images from the corresponding cropped regions of $C'$.

Theoretically, \textit{LDH can support up to $\omega^2$ secret images (i.e., $\omega^2 \times 24$ bpp) simultaneously} if the communication parties agree on the value of $\omega$ in advance. And different from existing multi-image hiding schemes (e.g., UDH \cite{zhang2020udh}), LDH can conceal multiple images using onle one set of $H$, $R$, and $P$, not to train multiple network pairs. Thus, our LDH is more efficient. The multiple image hiding in LDH can be represented as
\begin{equation}
    C'=Local\_Add \left (C,\{H(S^{(k)}, \omega)\} \right ),
\end{equation}
where $S^{(k)}$ is the $k$-th secret image. And the corresponding revealing process is
\begin{equation}
    \{S'^{(k)}\}=R(Cr(C', P(C'))).
\end{equation}


In addition to significantly increase the embedding rate, LDH has another advantage: LDH can further reduce the risk of being detected the existence of the sharing secrets by embedding $S_e$ within texture-rich local regions of $C$. For example, one can use the Just Noticeable Difference \cite{wu2017enhanced} (JND) technique to measure the human eye's sensitivity to distortion in different regions and find the most suitable positions for concealing secret images, which would obtain a $C'$ with better visual quality.

\section{Experiments}\label{sec:experimental_evaluation}

\subsection{Configurations}
\textbf{Dataset.} We use ImageNet and COCO to evaluate all involved hiding schemes \cite{baluja2017hiding,zhang2020udh,yu2020attention,lu2021large,guan2022deepmih}. For each dataset, we randomly select 10,000 images, including 8,000 for training, 1,000 for validation and 1,000 for testing. All selected images are resized to $1024\times1024$, and pixel values are normalized within $\left [ 0, 1 \right ]$.

\textbf{Implementation of LDH and Baselines.}
In the implementation of LDH, we first consider $\omega=4$ (i.e., the size of $S_e$ outputted by $H$ is $256\times256$), and we discuss other values of $\omega$ in later subsection. For $H$, we adopt a convolution layer and a max pooling layer to build its shallow layers for acquiring small-size features maps. Then, we adopt UNet \cite{ronneberger2015u} to build its remaining layers, which is responsible for generating small-size yet imperceptible secret codes, i.e., compact codes $S_e$. To produce the binary map, we design the network $P$ as a full-convolution network with six convolution layers, which keeps the size of its input and output consistent. As for the network $R$, we also use six convolution layers to build its shallow layers for preprocessing its inputs, and we take the network proposed in \cite{dong2016accelerating} to gradually increase the size of feature maps until meeting the size of the original secret images. As described in the previous section, in the pre-training phase, we set $\lambda_1=0.25$, $\lambda_2=0$, and $\lambda_3=0.75$ to obtain a well-trained $H$ and $R$. In the co-training phase, the main goal is to train $P$ that can accurately locate the position of $S_e$, so we set $\lambda_1=0.1$, $\lambda_2=0.8$ and $\lambda_3=0.1$. Besides, we set the initial learning rate to 0.001 and scale it down by a factor of 0.1 after every 30 epochs during the training process. All networks are trained with the Adam optimizer \cite{kingma2014adam}.

We also consider some baselines for comparison, including DDH \cite{baluja2017hiding}, UDH \cite{zhang2020udh}, HCVS \cite{weng2019high} and MISDNN \cite{das2021multi}. We follow the public-available implementations\footnote{https://github.com/arnoweng/PyTorch-Deep-Image-Steganography}$^,$\footnote{https://github.com/ChaoningZhang/Universal-Deep-Hiding}$^,$\footnote{https://github.com/muziyongshixin/pytorch-Deep-Steganography}$^,$\footnote{https://github.com/m607stars/MultiImageSteganography} to implement the four baselines respectively. We also achieve multiple-image hiding for DDH \cite{baluja2017hiding}, UDH \cite{zhang2020udh}, and HCVS \cite{weng2019high} by concatenating multiple secret images in the channel dimension. For MISDNN \cite{das2021multi}, we adjust the embedding rate by increasing or decreasing the number of network pairs.
\begin{table*}[t]
    \centering
    \caption{Performance comparison. $\uparrow$ denotes higher value is better, and vice versa.}
    \label{table:performance}
    \resizebox{0.95\linewidth}{!}{
        \begin{tabular}{|c|ccc|ccc|ccc|ccc|}
            \hline
            \multirow{3}{*}{Method}                         & \multicolumn{6}{c|}{Cover/Stogo 1 pair}                   & \multicolumn{6}{c|}{Secret/Recovery 1 pair}                \\
            \cline{2-13}
                                                            & \multicolumn{3}{c|}{ImageNet} & \multicolumn{3}{c|}{COCO} & \multicolumn{3}{c|}{ImageNet} & \multicolumn{3}{c|}{COCO}  \\
            \cline{2-13}
                                                            & APD$\downarrow$   & SSIM$\uparrow$   & PSNR$\uparrow$         & APD$\downarrow$   & SSIM$\uparrow$   & PSNR$\uparrow$     & APD$\downarrow$   & SSIM$\uparrow$   & PSNR$\uparrow$         & APD$\downarrow$   & SSIM$\uparrow$   & PSNR$\uparrow$      \\
            \hline
            DDH \cite{baluja2017hiding}     & 2.552  & 0.961 & 37.963       & 2.417  & 0.978 & 37.942   & 2.261  & 0.991  & 39.055      & 1.956  & 0.989 & 39.614    \\
            UDH \cite{zhang2020udh}         & 2.031  & 0.979 & 40.035       & 2.121  & 0.981 & 39.746   & 2.850  & 0.969  & 35.796      & 2.694  & 0.975 & 36.708    \\
            MISDNN \cite{das2021multi}      & 2.607  & 0.974 & 37.413       & 2.411  & 0.980 & 37.780   & 4.309  & 0.970  & 33.065      & 4.301  & 0.969 & 33.146    \\
            HCVS \cite{weng2019high} & 2.566  & 0.955 & 37.571       & 2.531  & 0.961 & 37.763   & 2.243  & 0.989  & 39.032      & 2.253  & 0.982 & 38.978    \\
            Ours                            & 0.346  & 0.994 & 43.887       & 0.292  & 0.996 & 45.118   & 3.809  & 0.951  & 33.169      & 3.974  & 0.947 & 32.963    \\
            \hline
            \hline
            \multirow{3}{*}{Method}         & \multicolumn{6}{c|}{Cover/Stogo 2 pairs}                   & \multicolumn{6}{c|}{Secret/Recovery 2 pairs}                \\
            \cline{2-13}
                                            & \multicolumn{3}{c|}{ImageNet} & \multicolumn{3}{c|}{COCO} & \multicolumn{3}{c|}{ImageNet} & \multicolumn{3}{c|}{COCO}  \\
            \cline{2-13}
                                            & APD$\downarrow$   & SSIM$\uparrow$   & PSNR$\uparrow$         & APD$\downarrow$   & SSIM$\uparrow$   & PSNR$\uparrow$     & APD$\downarrow$   & SSIM$\uparrow$   & PSNR$\uparrow$         & APD$\downarrow$   & SSIM$\uparrow$   & PSNR$\uparrow$      \\
            \hline
            DDH \cite{baluja2017hiding}     & 3.165  & 0.964  & 36.150      & 3.179  & 0.964 & 35.815   & 3.190  & 0.984  & 35.960      & 3.128  & 0.982 & 35.812    \\
            UDH \cite{zhang2020udh}         & 3.398  & 0.954  & 35.940      & 3.376  & 0.953 & 36.006   & 4.632  & 0.953  & 32.435      & 4.502  & 0.956 & 32.846    \\
            MISDNN \cite{das2021multi}      & 6.479  & 0.935  & 30.086      & 6.040  & 0.936 & 30.177   & 7.035  & 0.934  & 29.089      & 7.046  & 0.931 & 28.940    \\
            HCVS. \cite{weng2019high} & 3.688  & 0.953  & 34.839      & 3.808  & 0.953 & 34.331   & 3.467  & 0.977  & 35.412      & 3.517  & 0.968 & 35.092    \\
            Ours                            & 0.657  & 0.987  & 41.188      & 0.572  & 0.989 & 42.185   & 3.809  & 0.951  & 33.169      & 3.974  & 0.947 & 32.963    \\
            \hline
            \hline
            \multirow{3}{*}{Method}         & \multicolumn{6}{c|}{Cover/Stogo 3 pairs}                   & \multicolumn{6}{c|}{Secret/Recovery 3 pairs}                \\
            \cline{2-13}
                                            & \multicolumn{3}{c|}{ImageNet} & \multicolumn{3}{c|}{COCO} & \multicolumn{3}{c|}{ImageNet} & \multicolumn{3}{c|}{COCO}  \\
            \cline{2-13}
                                            & APD$\downarrow$   & SSIM$\uparrow$   & PSNR$\uparrow$         & APD$\downarrow$   & SSIM$\uparrow$   & PSNR$\uparrow$     & APD$\downarrow$    & SSIM$\uparrow$   & PSNR$\uparrow$        & APD$\downarrow$   & SSIM$\uparrow$   & PSNR$\uparrow$      \\
            \hline
            DDH \cite{baluja2017hiding}     & 4.160  & 0.940  & 33.824        & 4.270  & 0.937 & 33.641   & 3.806   & 0.978 & 34.391      & 3.846  & 0.978  & 34.340   \\
            UDH \cite{zhang2020udh}         & 4.120  & 0.933  & 34.392        & 4.145  & 0.935 & 34.336   & 6.430   & 0.936 & 30.321      & 6.110  & 0.932  & 30.375   \\
            MISDNN \cite{das2021multi}      & 11.102 & 0.886  & 25.694        & 9.647  & 0.893 & 26.407   & 10.708  & 0.892 & 25.682      & 10.17  & 0.903  & 25.957   \\
            HCVS \cite{weng2019high} & 5.080  & 0.912  & 32.301        & 5.056  & 0.930 & 32.177   & 3.802   & 0.971 & 34.616      & 3.871  & 0.965  & 34.130   \\
            Ours                            & 0.983  & 0.982  & 39.572        & 0.988  & 0.984 & 39.651   & 3.809   & 0.951 & 33.169      & 3.974  & 0.947  & 32.963   \\
            \hline
            \hline
            \multirow{3}{*}{Method}         & \multicolumn{6}{c|}{Cover/Stogo 4 pairs}                   & \multicolumn{6}{c|}{Secret/Recovery 4 pairs}                \\
            \cline{2-13}
                                            & \multicolumn{3}{c|}{ImageNet} & \multicolumn{3}{c|}{COCO} & \multicolumn{3}{c|}{ImageNet} & \multicolumn{3}{c|}{COCO}  \\
            \cline{2-13}
                                            & APD$\downarrow$   & SSIM$\uparrow$   & PSNR$\uparrow$         & APD$\downarrow$   & SSIM$\uparrow$   & PSNR$\uparrow$     & APD$\downarrow$    & SSIM$\uparrow$   & PSNR$\uparrow$        & APD$\downarrow$    & SSIM$\uparrow$   & PSNR$\uparrow$     \\
            \hline
            DDH \cite{baluja2017hiding}     & 5.758  & 0.927 & 31.118       & 5.611  & 0.923 & 31.147   & 6.679   & 0.964 & 29.909      & 6.440   & 0.962  & 30.061  \\
            UDH \cite{zhang2020udh}         & 4.740  & 0.915 & 33.163       & 4.755  & 0.917 & 33.139   & 10.785  & 0.891 & 25.725      & 10.414  & 0.882  & 25.767  \\
            MISDNN \cite{das2021multi}      & 16.38  & 0.823 & 21.797       & 18.38  & 0.817 & 20.862   & 13.667  & 0.841 & 23.243      & 12.117  & 0.855  & 24.095  \\
            HCVS \cite{weng2019high} & 5.076  & 0.919 & 31.948       & 5.331  & 0.921 & 31.803   & 6.806   & 0.947 & 29.505      & 6.817   & 0.955  & 29.990  \\
            Ours                            & 1.377  & 0.974 & 37.901       & 1.140  & 0.983 & 39.283   & 3.809   & 0.951 & 33.169      & 3.974   & 0.947  & 32.963  \\
            \hline
        \end{tabular}
    }
\end{table*}

\textbf{Evaluation Metrics.} To quantify the performance of all involved schemes, we use three visual metrics to measure the distance between $C$ and $C'$ (or $S$ and $S'$), including Average Pixel Discrepancy (APD), Peak Signal-to-Noise Ratio (PSNR), and Structural Similarity Index (SSIM).
\begin{itemize}
    \item \textbf{APD}. APD is the average $L_1$-norm distance between images, which can be formulated as
    \begin{equation}
        APD=\frac{1}{WH}\sum_{i=1}^{W}\sum_{j=1}^{H}\left | I_{i,j}-K_{i,j} \right |,
    \end{equation}
    where $I$ and $K$ are the reference image ($C$ or $S$) and the corresponding distorted image ($C'$ or $S'$) respectively. $H$ and $W$ are the height and width of images. The coordinate $(i,j)$ reflects the position of pixel $I_{i,j}$ (or $K_{i,j}$).

    \item \textbf{PSNR}. PSNR is a popular metric used to quantify the quality of signal reconstruction and is wildly used in related study. It is constructed based on $L_2$-norm as
    \begin{equation}
            PSNR = 10\cdot\log_{10}{\left ( \frac{MAX_I^2}{MSE} \right )},
    \end{equation}
    where
    \begin{equation}
        MSE = \frac{1}{WH}\sum_{i=1}^{W}\sum_{j=1}^{H}\left [ I_{i,j}-K_{i,j} \right ]^2,
    \end{equation}
    and $MAX_I$ indicates the maximum possible pixel value of the image, which is 1 in experiments.

    \item \textbf{SSIM}. SSIM measures the similarity between two images mainly based on the perceived change in structural information. It is also commonly used in related studies, which can be represented as
    \begin{equation}
        SSIM=\frac{(2\mu_I\mu_K+c_1)(2\sigma_{IK}+c_2)}{(\mu_I^2+\mu_K^2+c_1)(\sigma_I^2+\sigma_K^2+c_2)}
    \end{equation}
    where $\mu_I$ (or $\mu_K$) is the mean values of $I$ (or $K$), $\sigma_I$ (or $\sigma_K$) the variances of $I$ (or $K$), and $\sigma_{IK}$ is the covariance between $I$ and $K$. $c_1$ and $c_2$ are used to maintain stability, where $c_1=(k_1L)^2$ and $c_2=(k_2L)^2$. $L$ has the same meaning as $MAX_I$. By default, we set $k_1=0.01$ and $k_2=0.03$.    
\end{itemize}
The smaller the APD value, or the larger the PSNR and SSIM values, the better the quality of the generated image, meaning better hiding or revealing performance.




\begin{figure*}[t]
    \centering
    \includegraphics[width=0.95\linewidth]{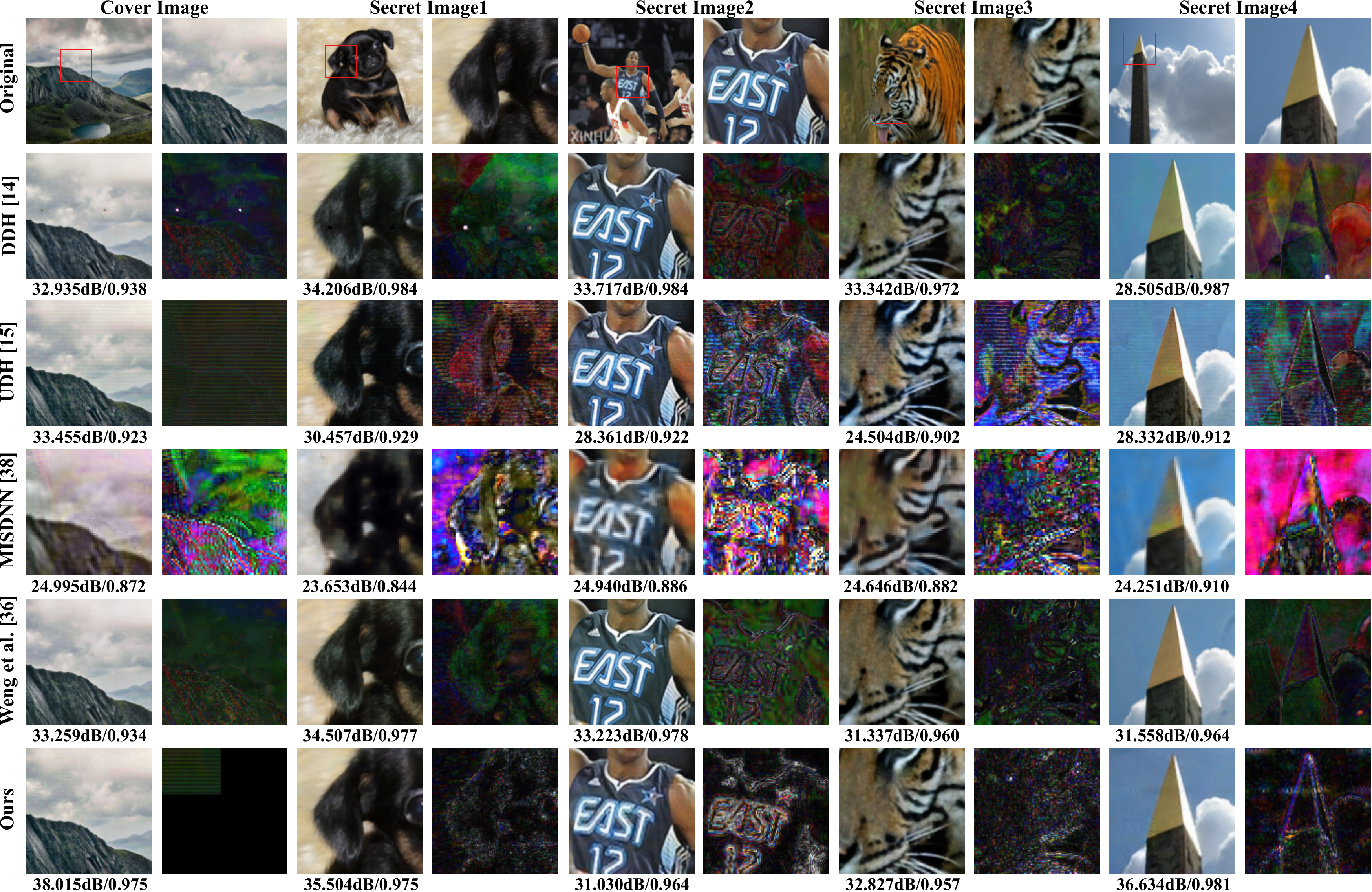}
    \caption{Visualization examples of hiding four secret images in one cover image. The first two images in the top row are the original cover image and the corresponding enlarged version of the region in red box, respectively. The remaining images in the top row are the revealed images and the enlarged local areas. From the second row to the bottom row, images in the odd columns are the enlarged local areas of the stego or revealed secret images produced by the marked hiding schemes, and images in the even columns are the difference images between the stego and cover images (or the revealed secret and original secret images). To facilitate observation, pixel values in each shown difference image are scaled by 7.}
    \label{fig:examples}
\end{figure*}

\subsection{Overall Evaluation}
\textbf{Quantitative Results.}
We evaluate all the involved schemes to hide 1, 2, 3 and 4 secret images within a cover image, and show the numerical results in Table \ref{table:performance}. From Table \ref{table:performance}, we can observe that our LDH performs much better than baselines in the hiding phase where the quality of $C'$ generated by LDH is best on all three metrics. With the compact secret codes, LDH can maintain the quality of $C'$ at a much high level (PNSR $>$ 37 dB) even after hiding up to 4 full-size secret images within a cover image.

For the revealing phase, LDH also shows its superiority. As one can observe from Table \ref{table:performance}, the quality of $S'$ does not be affected by the number of secret images in LDH. Instead, these baselines is not suitable for embedding more images because the quality of the secrets decreases as the number of the embedded secrets increase. The main reason for this gap between LDH and the baselines is that the hiding and revealing process of each secret image in LDH is independent. Thus, the revealing of different secret images will not affect each other, but the the multiple-image hiding of the baselines would cause mutual influence between channels. In the case of only hiding one secret image, the revealing quality of LDH is slightly lower than DDH \cite{baluja2017hiding}, UDH \cite{zhang2020udh} and HCVS \cite{weng2019high}, but similar with that of MISDNN \cite{das2021multi}. We think this is due to the fact that the compact secret codes cause the loss of some secret information compared with the full-size imperceptible secret codes.

\begin{figure}[t]
    \centering
    \includegraphics[width=1.0\linewidth]{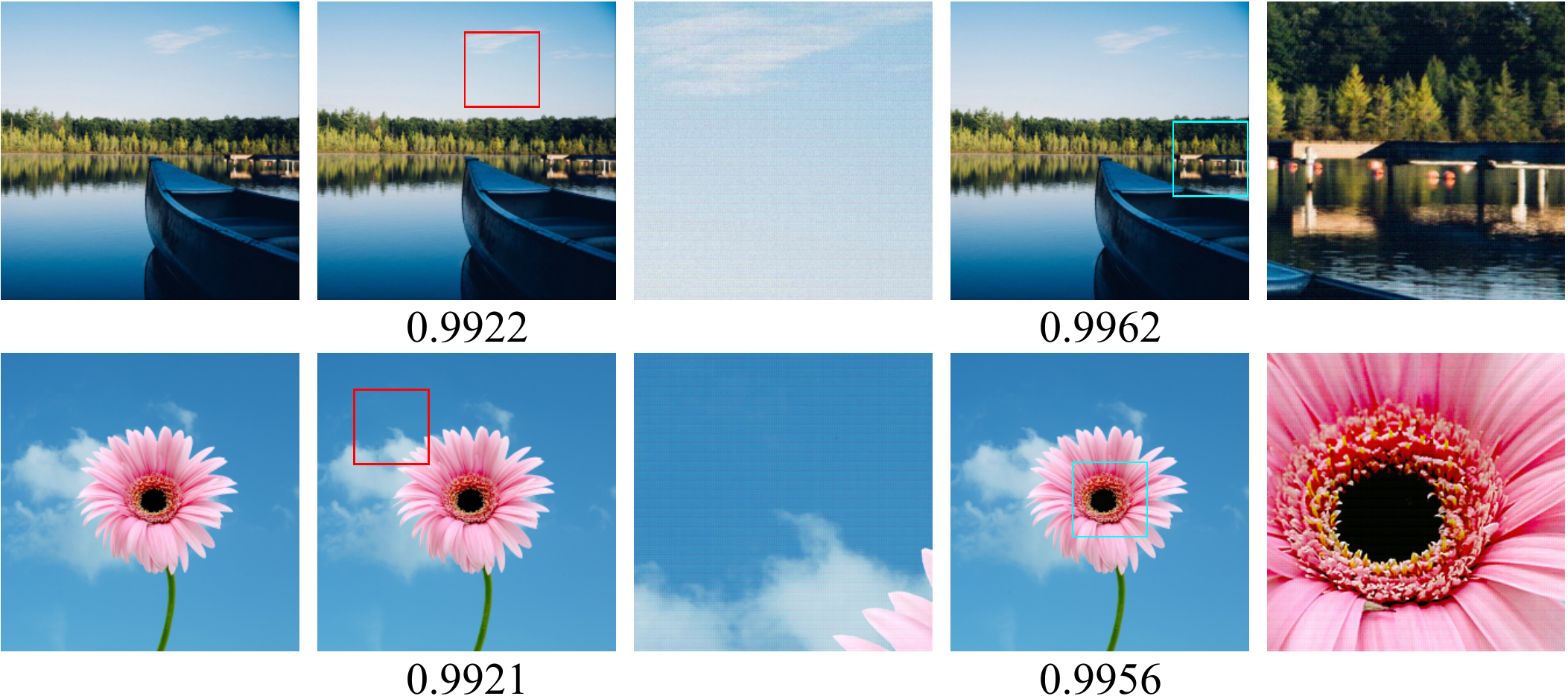}
    \caption{Visual comparison of JND-guided position versus random position embedding the secret codes $S_e$. \textbf{Column 1:} The original cover image $C$. \textbf{Column 2:} The stego image $C'$ generated by random position embedding: $S_e$ is embedded into the red boxes. \textbf{Column 3:} Enlarge the parts in the red boxes. \textbf{Column 4:} The $C'$ generated under the guidance of JND: $S_e$ is embedded into the blue boxes. \textbf{Column 5:} Enlarge the parts in the blue boxes. The numbers below $C'$ indicates the SSIM value compared with $C$.}
    \label{fig:jnd}
\end{figure}

\textbf{Qualitative Results.} We exhibit some visualization results in Fig. \ref{fig:examples} to further show the superiority of our LDH. First, in the hiding phase, LDH brings the smallest perturbations for the cover images (see the column 2 in Fig. \ref{fig:examples}). The baselines are of global hiding, where the perturbations in the smooth area of the stego image can be easily detected, such as the perturbations in the sky. Specifically, DDH \cite{baluja2017hiding} may produce noise points in local areas; UDH \cite{zhang2020udh} generates a stego image with undesired horizontal stripes; MISDNN \cite{das2021multi} produces severe color distortion; and HCVS \cite{weng2019high} has obvious distortion at the edges of the secret images. Although LDH still produces horizontal stripes, we can avoid these stripes being easily found by selecting textured local areas for hiding. For example, in Fig. \ref{fig:jnd}, we use JND method \cite{wu2017enhanced} to determine more suitable local area (e.g., the woods) for carrying secret information, which achieves better quality of the stego images both subjectively and objectively.

\begin{figure}[t]
    \centering
    \subfigure[]{\includegraphics[width=0.48\linewidth]{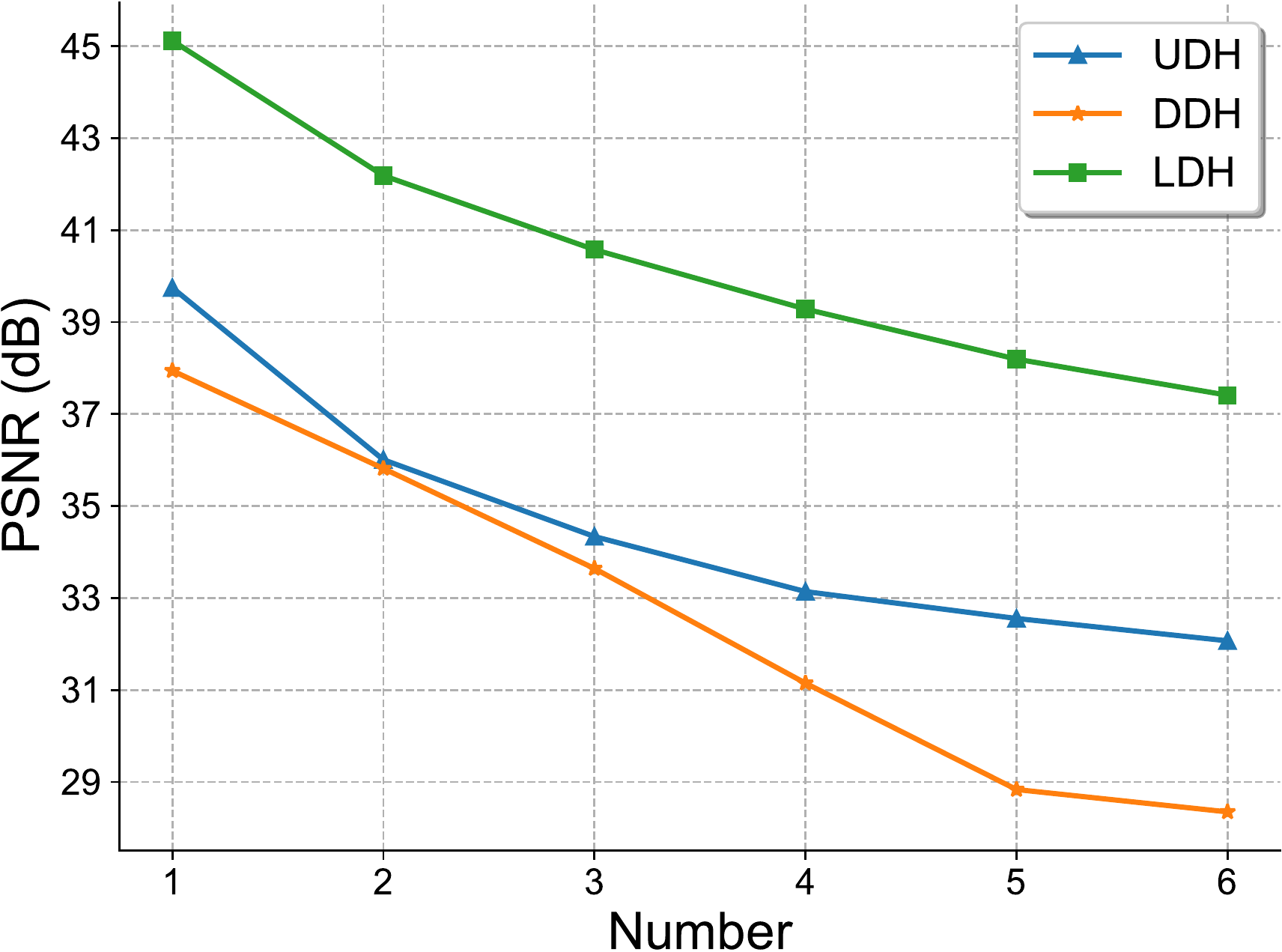}\label{fig:stego_quality:a}}
    \hfill
    \subfigure[]{\includegraphics[width=0.48\linewidth]{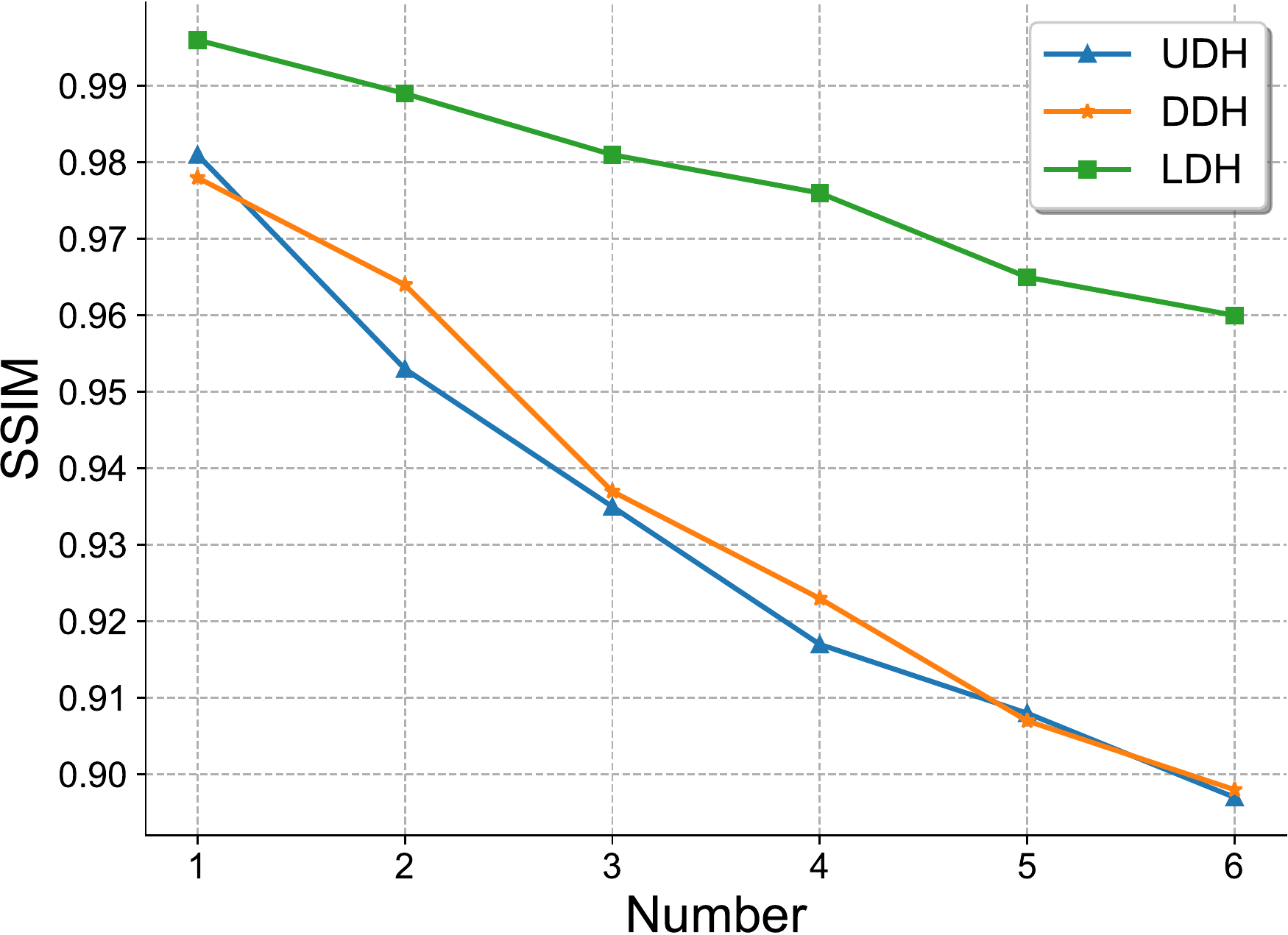}\label{fig:stego_quality:b}}
    \\
    \subfigure[]{\includegraphics[width=0.48\linewidth]{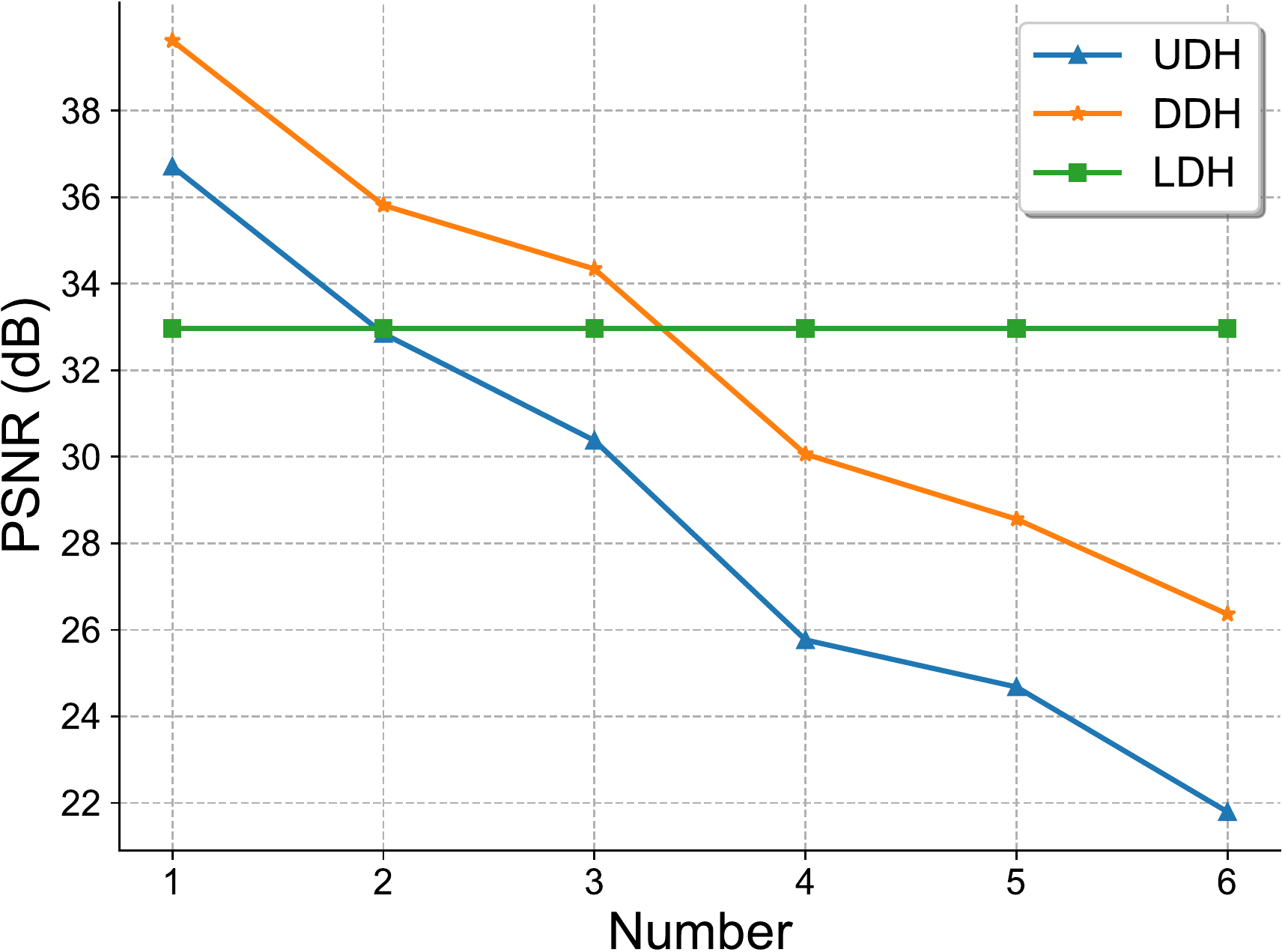}\label{fig:stego_quality:c}}
    \hfill
    \subfigure[]{\includegraphics[width=0.48\linewidth]{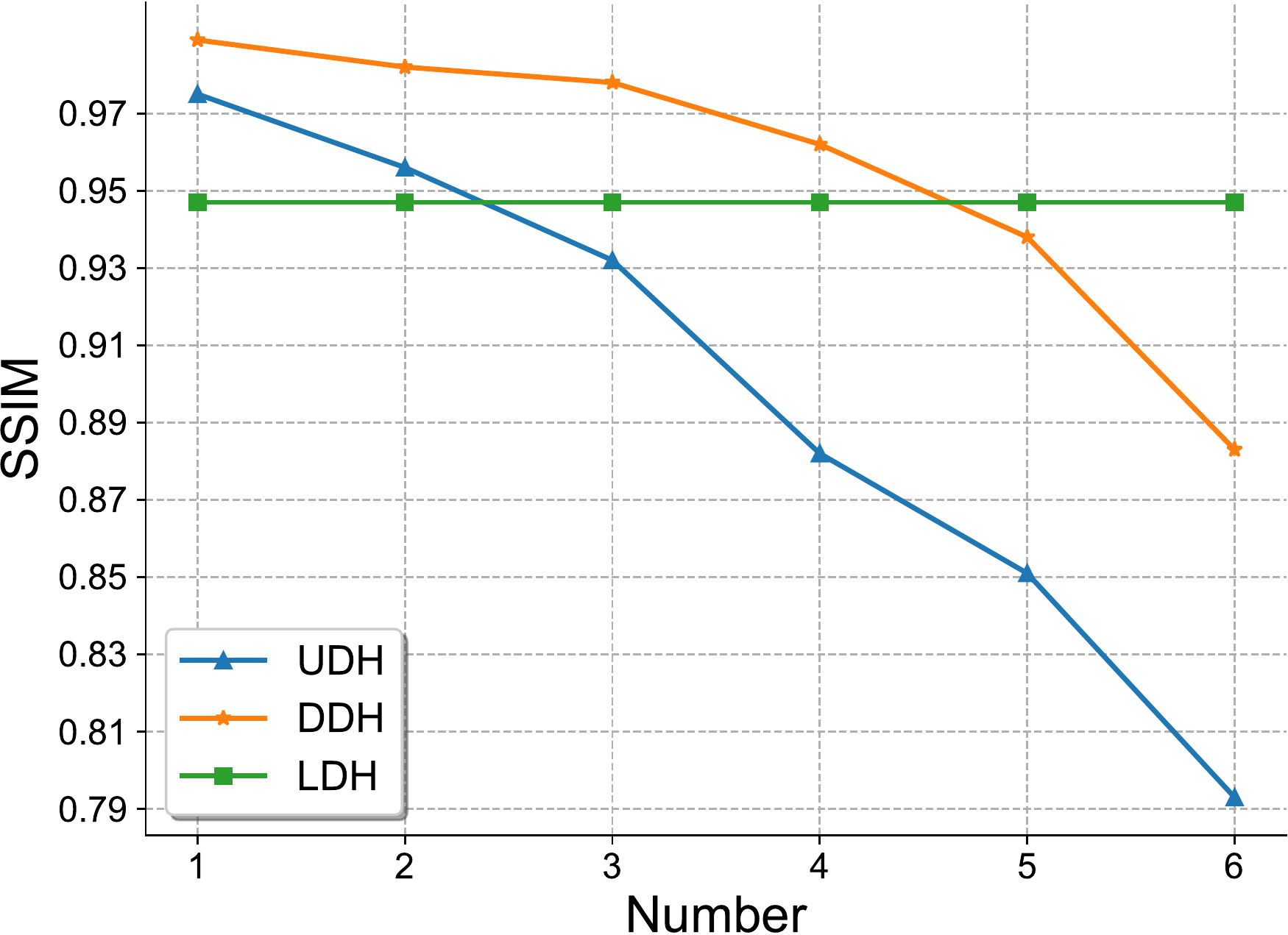}\label{fig:stego_quality:d}}

    \caption{Visual results of UDH, DDh, and LDH with different number of embedded images. (a) PSNR values of the cover/stego image pairs; (b) SSIM values of the cover/stego image pairs; (c) PSNR values of the secret/recovered image pairs; (d) SSIM values of the secret/recovered image pairs.}
    \label{fig:stego_quality}
\end{figure}

In the revealing phase, LDH still performs satisfactory. As shown in Fig. \ref{fig:examples}, we hide four different secret images in one cover image. From the revealed secret images we can see that LDH realizes the highest visual quality. By contrast, artifacts and colour distortions appear in the secret images revealed by other schemes. And these undesired distortions will become more seriously as the number of secret images increases (see Fig. \ref{fig:stego_quality}). Fortunately, our LDH is not bothered by this problem due to the characteristic of the local embedding.

We also illustrate the image quality comparison results of DDH \cite{baluja2017hiding}, UDH \cite{zhang2020udh} and our LDH under different embedding rates (from 1$\times$24 bpp to 6$\times$24 bpp) in Fig. \ref{fig:stego_quality} . The high quality of stego images is critical since it can better avoid manual or machine censorship. From Fig. \ref{fig:stego_quality:a} and \ref{fig:stego_quality:b},
we can observe that for the baselines, the quality of the stego image gradually declines as the number of embedded images increases since more perturbations are introduced into the cover image to embed more images. Instead, our LDH still performs better at maintaining the quality of stego images, especially when hiding multiple secret images. In addition to the quality of stego images, Fig. \ref{fig:stego_quality:c} and \ref{fig:stego_quality:d} further show that in LDH, the quality of revealed secret images does not influenced by the number of embedded images. However, the corresponding visual quality of the baselines decreases seriously. From the whole Fig. \ref{fig:stego_quality}, we can conclude that LDH is much more suitable for high-capacity data hiding than these global ones.

\begin{figure}[t]
    \centering
    \includegraphics[width=0.9\linewidth]{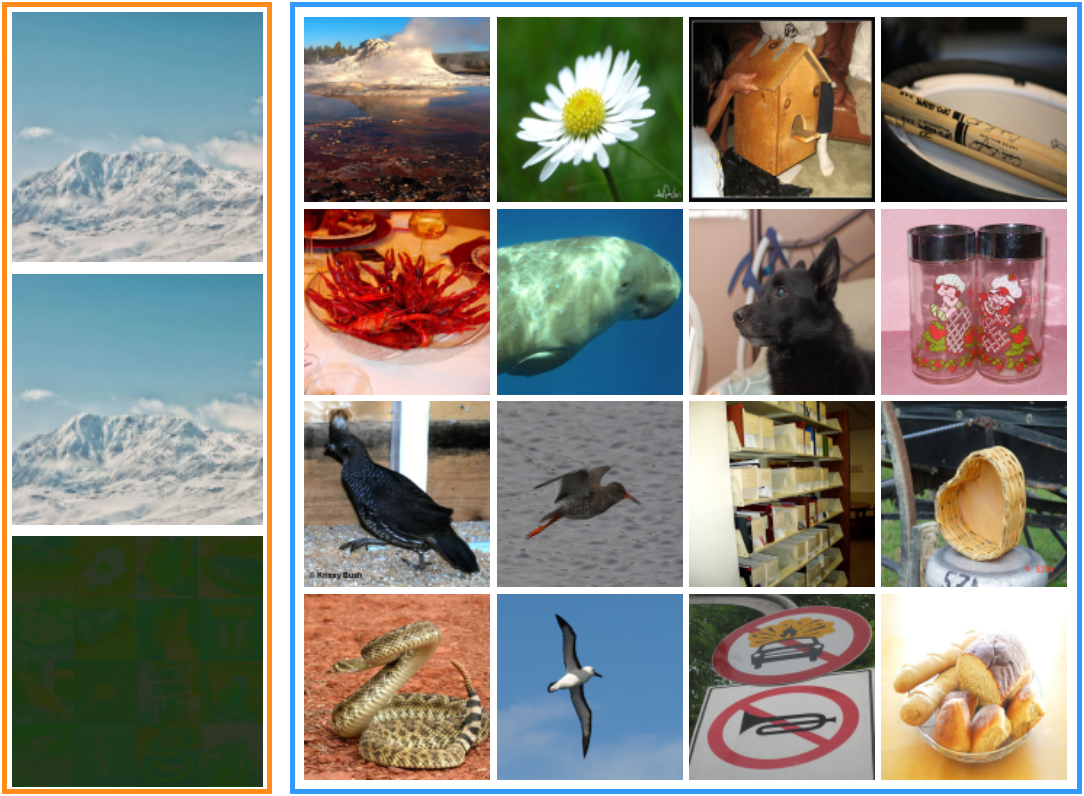}
    \caption{Illustration of the maximum embedding rate (i.e., 16$\times$24 bpp) of LDH when $\omega$ is 4. Images in the yellow box are $C$, $C'$, $\times$15 magnified errors between $C'$ and $C$ from top to bottom. The 16 revealed images $S'$ are in the blue box.}
    \label{fig:full_embed}
\end{figure}

\subsection{Impact of Hyper-parameters}
\textbf{Scaling Factor.} In our LDH, we use a scaling factor $\omega$ to determine the size of the compact secret codes, which also determines the upper bound of the embedding rate.
A larger  $\omega$ indicates a higher embedding rate.
However, a large $\omega$ also means extremely compact secret codes, which inevitably results in the loss of secret information and makes the recover of  high-quality secret images difficult. If the quality of the revealed secret image is too low, the corresponding hiding is meaningless and failed. We confirm this phenomenon in Table \ref{table:diff_scale}.
From Table \ref{table:diff_scale} and the results of LDH in Table \ref{table:performance}, we find that $\omega=4$ can balance the embedding rate and the quality of $S'$ better. As $\omega$ increases from 2 to 4, the quality of stego images is improved and the embedding rate reaches $16\times24$ bpp while sacrificing the quality of $S'$ about 6.24\%. The quality of $S'$ is acceptable when $\omega=4$ and the PSNR between the original and the revealed secret images is greater than 32 dB. Fig. \ref{fig:full_embed} illustrates the visual results of our LDH under $\omega=4$, where the revealing results of our LDH is almost the same as the revealing results when only embedding one image, i.e., all the recovered images are of high quality.
As we further increase $\omega$ from 4 to 8, the quality of $S'$ becomes unacceptable (the corresponding PSNR is only about 26 dB) although the theoretical embedding rate can be $64\times24$ bpp.

\begin{table*}[t]
    \centering
    \caption{The performance of our LDH under different scaling factor $\omega$.}
    \label{table:diff_scale}
    \resizebox*{0.9\textwidth}{!}{
    \begin{tabular}{|c|c|c|c|c|c|c|c|c|c|c|c|c|}
        \hline
        \multirow{3}{*}{\textbf{Dataset}} & \multicolumn{6}{c|}{$\omega$\textbf{=2, 1 image}}                                                                      & \multicolumn{6}{c|}{$\omega$\textbf{=8, 1 image}}                                                                       \\
        \cline{2-13}
                                 & \multicolumn{3}{c|}{\textbf{Cover/Stego pair}}                   & \multicolumn{3}{c|}{\textbf{Secret/Recovery pair}}               & \multicolumn{3}{c|}{\textbf{Cover/Stego pair}}                   & \multicolumn{3}{c|}{\textbf{Secret/Recovery pair}}                \\
        \cline{2-13}
                                 & APD$\downarrow$ & SSIM$\uparrow$ & PSNR$\uparrow$ & APD$\downarrow$ & SSIM$\uparrow$ & PSNR$\uparrow$ & APD$\downarrow$ & SSIM$\uparrow$ & PSNR$\uparrow$ & APD$\downarrow$ & SSIM$\uparrow$ & PSNR$\uparrow$  \\
        \hline
        COCO                     & 1.321             & 0.976            & 38.246           & 3.527             & 0.965            & 33.93            & 0.114             & 0.998            & 48.19            & 8.606             & 0.822            & 25.891            \\
        \hline
        ImageNet                 & 1.36              & 0.972            & 38.078           & 2.676             & 0.977            & 36.33            & 0.117             & 0.998            & 48.045           & 8.078             & 0.843            & 26.665            \\
        \hline
        \multirow{3}{*}{\textbf{Dataset}} & \multicolumn{6}{c|}{$\omega$\textbf{=2, 2 images}}                                                                     & \multicolumn{6}{c|}{$\omega$\textbf{=8, 2 images}}                                                                      \\
        \cline{2-13}
                                 & \multicolumn{3}{c|}{\textbf{Cover/Stego pair}}                   & \multicolumn{3}{c|}{\textbf{Secret/Recovery pair}}               & \multicolumn{3}{c|}{\textbf{Cover/Stego pair}}                   & \multicolumn{3}{c|}{\textbf{Secret/Recovery pair}}                \\
        \cline{2-13}
                                 & APD$\downarrow$ & SSIM$\uparrow$ & PSNR$\uparrow$ & APD$\downarrow$ & SSIM$\uparrow$ & PSNR$\uparrow$ & APD$\downarrow$ & SSIM$\uparrow$ & PSNR$\uparrow$ & APD$\downarrow$ & SSIM$\uparrow$ & PSNR$\uparrow$  \\
        \hline
        COCO                     & 2.608             & 0.947            & 35.288           & 3.527             & 0.965            & 33.93            & 0.228             & 0.996            & 45.182           & 8.606             & 0.822            & 25.891            \\
        \hline
        ImageNet                 & 2.666             & 0.949            & 35.186           & 2.676             & 0.977            & 36.33            & 0.234             & 0.996            & 45.033           & 8.078             & 0.843            & 26.665            \\
        \hline
        \multirow{3}{*}{\textbf{Dataset}} & \multicolumn{6}{c|}{$\omega$\textbf{=2, 4 images}}                                                                     & \multicolumn{6}{c|}{$\omega$\textbf{=8, 4 images}}                                                                      \\
        \cline{2-13}
                                 & \multicolumn{3}{c|}{\textbf{Cover/Stego pair}}                   & \multicolumn{3}{c|}{\textbf{Secret/Recovery pair}}               & \multicolumn{3}{c|}{\textbf{Cover/Stego pair}}                   & \multicolumn{3}{c|}{\textbf{Secret/Recovery pair}}                \\
        \cline{2-13}
                                 & APD$\downarrow$ & SSIM$\uparrow$ & PSNR$\uparrow$ & APD$\downarrow$ & SSIM$\uparrow$ & PSNR$\uparrow$ & APD$\downarrow$ & SSIM$\uparrow$ & PSNR$\uparrow$ & APD$\downarrow$ & SSIM$\uparrow$ & PSNR$\uparrow$  \\
        \hline
        COCO                     & 5.34              & 0.88             & 32.141           & 3.527             & 0.965            & 33.93            & 0.454             & 0.992            & 42.177           & 8.606             & 0.822            & 25.891            \\
        \hline
        ImageNet                 & 5.381             & 0.889            & 32.118           & 2.676             & 0.977            & 36.33            & 0.468             & 0.991            & 42.028           & 8.078             & 0.843            & 26.665            \\
        \hline
    \end{tabular}}
\end{table*}

\begin{table}[t]
    \centering
    \caption{Image quality Results of the revealed images under different capacities of the revealing networks. The results are measured by the PSNR and SSIM values between the secret and revealed images.}
    \label{table:decoder}
    \begin{tabular}{@{}cccc@{}}
        \toprule
        Models & nhf=48 & nhf=64 & nhf=80\\
        \midrule
        PSNR & 31.382 & 32.537 & 33.169\\
        SSIM & 0.930 & 0.942 & 0.951\\
        \bottomrule
    \end{tabular}
\end{table}

\textbf{Model Capacity.} In deep hiding, DNNs have a powerful representation ability to encode image information into their parameters, which is the biggest advantage of deep hiding compared to traditional data hiding and is the key to achieve high embedding rate. Besides, we think that deep neural networks with a higher capacity can support a higher embedding rate. To confirm it, we test multiple networks of different scales. Specifically, we change the width of all layers in the revealing network to control the number of hidden features (nhf), and record the corresponding experimental results in Table \ref{table:decoder}. As we can observe from Table \ref{table:decoder}, a larger nhf (i.e., a wider revealing network) leads to higher PSNR and SSIM scores. Therefore, model capacity is critical for a high embedding rate in deep hiding.


\subsection{Robustness Analysis}
During the transmission of $C'$, $C'$ could be distorted by unavoidable channel noises or artificial perturbations (e.g., JPEG compression, rotation, and cropping). In this case, strong robustness allows hiding schemes more practical in the real world.
Next, we will show that our LDH can naturally resist local distortions, and can be further enhanced by adversarial training to resist global distortions.


\textbf{Local Distortions.} Due to the special local embedding strategy, LDH can resist local distortions, e.g., random crop. Existing global deep hiding schemes suffer from the weakness of low redundancy \cite{xiang2021peel} and the missing of any part of a stego image will lead to wrong recovery of the corresponding secret pixels.
However, the missing of a random part of the stego image in our LDH may not affect the revealing of the secret image since the secret code is embedded locally and randomly. Thus, it is still possible for LDH to reveal the complete secret image after the stego image has experienced some local distortions. And the more compact of the secret code (i.e., larger $\omega$), the higher the probability of the complete recovery is.

\begin{figure*}[!t]
    \centering
    \resizebox*{\textwidth}{!}{
        \huge
        \begin{tabular}{ccccccccc}
            & Stego image                                                                             & Global schemes                                                                            & Ours-1 & Ours-2                                                                                       & Stego image                                                                             & Other                                                                                       & Ours-1 & Ours-2                                                                                         \\
    \begin{sideways}Cropout\end{sideways} & \makecell*[c]{\includegraphics[width=0.30\linewidth]{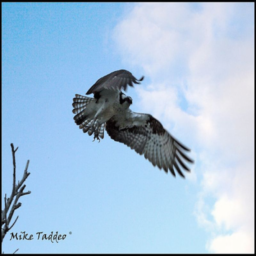}} & \makecell*[c]{\includegraphics[width=0.30\linewidth]{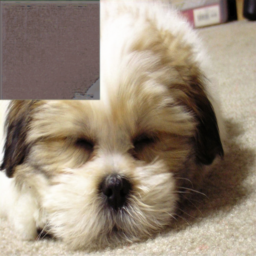}} & \makecell*[c]{\includegraphics[width=0.30\linewidth]{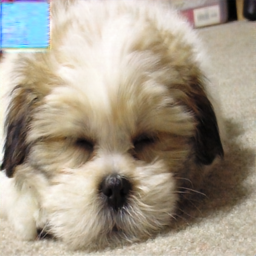}} & \makecell*[c]{\includegraphics[width=0.30\linewidth]{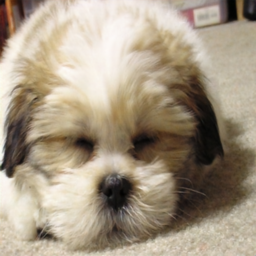}} & \makecell*[c]{\includegraphics[width=0.30\linewidth]{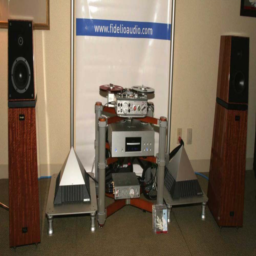}} & \makecell*[c]{\includegraphics[width=0.30\linewidth]{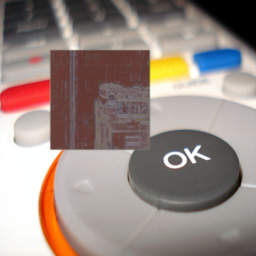}} & \makecell*[c]{\includegraphics[width=0.30\linewidth]{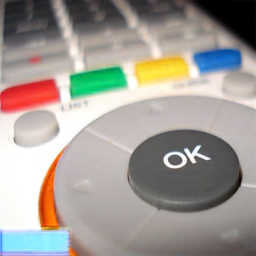}} & \makecell*[c]{\includegraphics[width=0.30\linewidth]{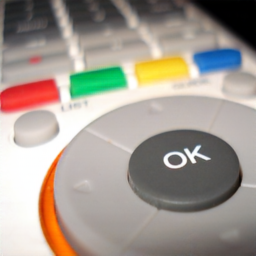}}  \\
    \begin{sideways}Crop\end{sideways}    & \makecell*[c]{\includegraphics[width=0.30\linewidth]{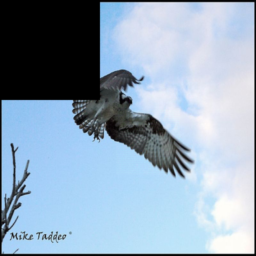}}    & \makecell*[c]{\includegraphics[width=0.30\linewidth]{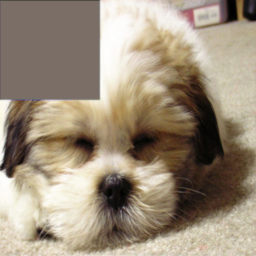}}    & \makecell*[c]{\includegraphics[width=0.30\linewidth]{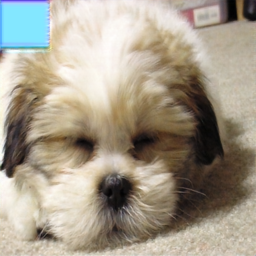}} & \makecell*[c]{\includegraphics[width=0.30\linewidth]{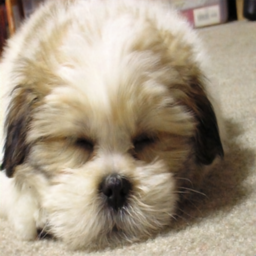}}    & \makecell*[c]{\includegraphics[width=0.30\linewidth]{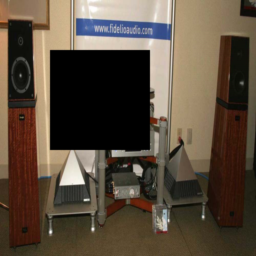}}    & \makecell*[c]{\includegraphics[width=0.30\linewidth]{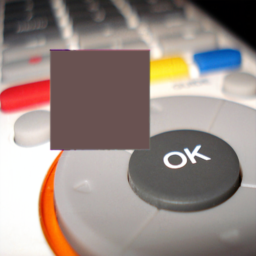}}    & \makecell*[c]{\includegraphics[width=0.30\linewidth]{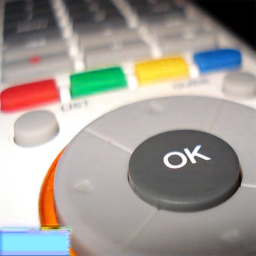}} & \makecell*[c]{\includegraphics[width=0.30\linewidth]{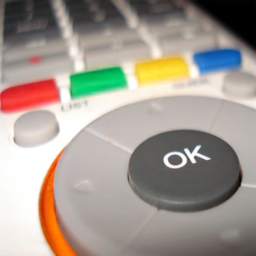}}
    \end{tabular}
    }
    \caption{Visualization of the recovery effect after Cropout and Crop attack on the stego image. The first four columns are attacked at the top left corner and the last four columns are attacked at a random position of the stego image. The size of the image is 1024$\times$1024 pixels and the size of the cropped region is 400$\times$400 pixels.}
    \label{fig:crop}
\end{figure*}

To show the innate robustness of LDH, we adopt a cropping operation to erase part pixels from stego images. In particular, after the cropping, we can fill the hole in the stego image with the corresponding part of the cover image (\textit{Cropout}), or directly fill the hole with 0 (\textit{Crop}). In Fig. \ref{fig:crop}, we use Crop and Cropout to process the stego images (the size of the cropped hole is $400\times400$). From Fig. \ref{fig:crop}, we can see the innate robustness of LDH against Crop and Cropout intuitively. The Crop and Cropout operations do not affect the revealing of the secret code in LDH if the cropped area and embedded area do not overlap. However, the global hiding schemes (e.g.,DDH \cite{baluja2017hiding} and UDH \cite{zhang2020udh}) are certainly unable to recover the complete $S'$ after the cropping operation.
To better illustrate this case, we also show the examples of the overlapping (column 3 and column 7) in Fig. \ref{fig:crop}. The specific numerical results of PSNR after the cropping attacks are exhibited in Table \ref{table:crop}, in which we first divide the stego image into 16 disjoint blocks and each block is 256$\times$256 pixels, and then randomly select 1, 2 and 4 blocks for executing Crop or Cropout. From Table \ref{table:crop}, it is clear to observe that LDH is more robust to cropping attacks than the other two global schemes due to the much higher quality of the revealed secret images.

\begin{table}[t]
    \centering
    \caption{The recovery quality results of DDH, UDH and our LDH after the suffering crop attacks. The results are measured by PSNR. The suffix number is the number of randomly cropped blocks.}
    \label{table:crop}
    \begin{tabular}{cccc}
        \toprule
        Method    & DDH    & UDH    & LDH     \\
        \hline
        Crop-1    & 23.699 & 21.573 & 30.863  \\
        Cropout-1 & 23.571 & 21.557 & 30.656  \\
        Crop-2    & 20.228 & 18.918 & 29.022  \\
        Cropout-2 & 20.217 & 18.814 & 29.843  \\
        Crop-4    & 17.225 & 16.509 & 26.059  \\
        Cropout-4 & 17.227 & 16.461 & 26.255  \\
        \bottomrule
    \end{tabular}
\end{table}

\textbf{Global Distortions.} Besides the local distortions, our LDH can also resist other common global distortions by targeted adversarial training \cite{zhang2020udh}. In the adversarial training, we add a noise layer between the hiding and revealing modules, which simulates possible distortions and processes $C'$. We then train $P$ and $R$ to identify the embedding regions and reveal desired secret images from the processed $C'$ respectively. In particular, we consider three different distortions here to verify the acquired enhancement of LDH: \textit{Dropout} (Dropout replaces each pixel of $C'$ with the corresponding pixel of $C$ with a certain probability. We set the probability to 0.3), \textit{Gaussian Blur} (short as Gaussian), and \textit{JPEG Compression} (short as JPEG). Then we use a Dropout layer, a Gaussian layer, and a JPEG layer to play the role of the noise layer in the adversarial training. Note that, we can only use one of the three noise layers to process $C'$ (specialized) or use all three layers to process $C'$ (combined).

We evaluate the robustness of the vanilla models (i.e., no adversarial training) and the corresponding enhanced models, and show the results in Table \ref{table:robustness}. From Table \ref{table:robustness}, it is clear to observe that before the adversarial training, the vanilla model does not work facing all considered distortions since the quality of the revealed secret images is extremely low (PNSR $<$ 13 dB), which means the corresponding secret images are useless. However, after the adversarial training, the robustness of LDH has been significantly improved. For the specialized models, the PSNR of the revealed secret images under all three distortions is 2--3 times better than the vanilla model. In addition, the quality of the revealed secret images obtained by the combined model is slightly lower than that of the specialized models, but still at an acceptable level. The results of the combined model show a significant improvement over the vanilla model, with robustness to multiple distortions and stronger generalization ability.

\begin{table*}[ht]
    \centering
    \caption{The visual quality results of stego images and secret images under different image distortions. ``Vanilla training'': training without distortions; ``Specialized training'': training with a single corresponding distortion; ``Combined training'': training with combined distortions. ``Original'' indicates that no image distortions are added during the experiments.}
    \label{table:robustness}
    \resizebox*{0.75\linewidth}{!}{
    \begin{tabular}{|c|ccccccccc|}
    \hline
    \multirow{3}{*}{\textbf{Distortion type}} & \multicolumn{9}{c|}{Stego}                                                                                                                                                                                                                      \\ \cline{2-10}
                                     & \multicolumn{3}{c|}{\textbf{Vanilla Model}}                                                         & \multicolumn{3}{c|}{\textbf{Specialized Models}}                                                   & \multicolumn{3}{c|}{\textbf{Combined Model}}                               \\ \cline{2-10}
                                     & \multicolumn{1}{c|}{APD$\downarrow$ }    & \multicolumn{1}{c|}{SSIM$\uparrow$ }  & \multicolumn{1}{c|}{PSNR$\uparrow$ }   & \multicolumn{1}{c|}{APD$\downarrow$ }    & \multicolumn{1}{c|}{SSIM$\uparrow$ }  & \multicolumn{1}{c|}{PSNR$\uparrow$ }   & \multicolumn{1}{c|}{APD$\downarrow$ }    & \multicolumn{1}{c|}{SSIM$\uparrow$ }  & PSNR$\uparrow$    \\ \hline
    Original                           & \multicolumn{1}{c|}{0.343}  & \multicolumn{1}{c|}{0.994} & \multicolumn{1}{c|}{43.928} & \multicolumn{1}{c|}{0.418}  & \multicolumn{1}{c|}{0.991} & \multicolumn{1}{c|}{42.766} & \multicolumn{1}{c|}{0.583}  & \multicolumn{1}{c|}{0.985} & 40.619 \\ \hline
    Dropout                     & \multicolumn{1}{c|}{0.336}  & \multicolumn{1}{c|}{0.993} & \multicolumn{1}{c|}{44.039} & \multicolumn{1}{c|}{0.413}  & \multicolumn{1}{c|}{0.991} & \multicolumn{1}{c|}{42.876} & \multicolumn{1}{c|}{0.588}  & \multicolumn{1}{c|}{0.987} & 40.562 \\ \hline
    Gaussian                    & \multicolumn{1}{c|}{0.344}  & \multicolumn{1}{c|}{0.993} & \multicolumn{1}{c|}{43.926} & \multicolumn{1}{c|}{0.474}  & \multicolumn{1}{c|}{0.990}  & \multicolumn{1}{c|}{41.878} & \multicolumn{1}{c|}{0.586}  & \multicolumn{1}{c|}{0.986} & 40.586 \\ \hline
    JPEG                        & \multicolumn{1}{c|}{0.344}  & \multicolumn{1}{c|}{0.994} & \multicolumn{1}{c|}{43.938} & \multicolumn{1}{c|}{0.557}  & \multicolumn{1}{c|}{0.987} & \multicolumn{1}{c|}{40.909} & \multicolumn{1}{c|}{0.585}  & \multicolumn{1}{c|}{0.986} & 40.597 \\ \hline
    \multirow{3}{*}{\textbf{Distortion type}} & \multicolumn{9}{c|}{Secret}                                                                                                                                                                                                                         \\ \cline{2-10}
                                     & \multicolumn{3}{c|}{\textbf{Vanilla Model}}                                                         & \multicolumn{3}{c|}{\textbf{Specialized Models}}                                                   & \multicolumn{3}{c|}{\textbf{Combined Model}}                               \\ \cline{2-10}
                                     & \multicolumn{1}{c|}{APD$\downarrow$ }    & \multicolumn{1}{c|}{SSIM$\uparrow$ }  & \multicolumn{1}{c|}{PSNR$\uparrow$ }   & \multicolumn{1}{c|}{APD$\downarrow$ }    & \multicolumn{1}{c|}{SSIM$\uparrow$ }  & \multicolumn{1}{c|}{PSNR$\uparrow$ }   & \multicolumn{1}{c|}{APD$\downarrow$ }    & \multicolumn{1}{c|}{SSIM$\uparrow$ }  & PSNR$\uparrow$    \\ \hline
    Original                           & \multicolumn{1}{c|}{3.667}  & \multicolumn{1}{c|}{0.957} & \multicolumn{1}{c|}{33.615} & \multicolumn{1}{c|}{4.477}  & \multicolumn{1}{c|}{0.934} & \multicolumn{1}{c|}{31.961} & \multicolumn{1}{c|}{9.960}   & \multicolumn{1}{c|}{0.834} & 25.431 \\ \hline
    Dropout                     & \multicolumn{1}{c|}{48.370}  & \multicolumn{1}{c|}{0.248} & \multicolumn{1}{c|}{12.420}  & \multicolumn{1}{c|}{7.564}  & \multicolumn{1}{c|}{0.885} & \multicolumn{1}{c|}{26.947} & \multicolumn{1}{c|}{11.685} & \multicolumn{1}{c|}{0.832} & 23.230  \\ \hline
    Gaussian                    & \multicolumn{1}{c|}{52.714} & \multicolumn{1}{c|}{0.309} & \multicolumn{1}{c|}{11.854} & \multicolumn{1}{c|}{5.047}   & \multicolumn{1}{c|}{0.917} & \multicolumn{1}{c|}{30.533} & \multicolumn{1}{c|}{9.787}  & \multicolumn{1}{c|}{0.856} & 24.927 \\ \hline
    JPEG                        & \multicolumn{1}{c|}{98.625} & \multicolumn{1}{c|}{0.187} & \multicolumn{1}{c|}{6.659}  & \multicolumn{1}{c|}{17.974} & \multicolumn{1}{c|}{0.737} & \multicolumn{1}{c|}{20.447} & \multicolumn{1}{c|}{16.956} & \multicolumn{1}{c|}{0.740}  & 20.711 \\ \hline
    \end{tabular}
    }
\end{table*}


\section{Embedding Rate Analysis}
\label{sec:embedding_analysis}
Data hiding can be viewed as a 2-player game, in which the sender embeds secret messages into a cover image, and the receiver recovers the secret messages from the stego image. As we have discussed in Table \ref{table:diff_scale}, the quality of revealed images is crucial in data hiding since serious distortions make the hiding meaningless. In this section, we discuss the embedding rate from the perspective of the receiver.


\begin{figure}[t]
    \centering
    \includegraphics[width=0.9\columnwidth]{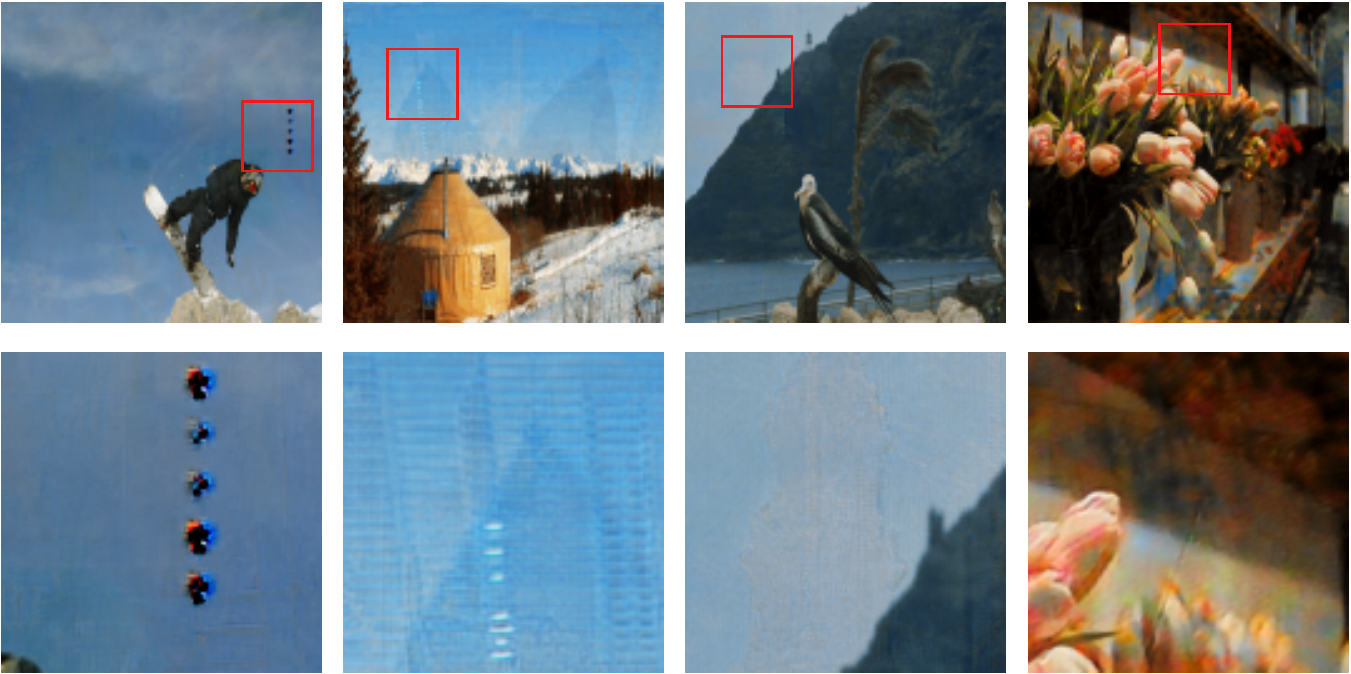}
    \caption{Recovered examples of DDH \cite{baluja2017hiding}, UDH \cite{zhang2020udh}, HCVS \cite{weng2019high} and MISDNN \cite{das2021multi} (from left to right) when hiding multiple images. The first row shows the revealed secret images and the second row shows the enlarged version of the regions in the boxes.}
    \label{fig:udh_ddh_multi}
\end{figure}

During the evaluation experiments, we find that the distortions on the recovery secret images (see Fig. \ref{fig:udh_ddh_multi}) are pattern-specific such as undesired stripe textures, the outline textures of $C$, and color distortion. These distortions may be reduced or even eliminated using extra image restoration algorithms. Therefore, we consider various receivers with different capacities to discuss the upper bound of the embedding rate of different deep hiding schemes. Specifically, we consider that the receiver can use non-learning-based algorithms or learning-based algorithms to further process the revealed secret images. Besides, it is necessary to set a threshold for the quality of recovered secret images to discuss the embedding rate. We set the the final embedding rate of a deep hiding scheme is the highest embedding rate that can be achieved when a specific scheme can meet this threshold.



\textbf{Receiver with Non-learning-based Algorithms.} We assume that the receiver can take some non-learning-based image enhancement or denoising algorithms to further process the recovered images here. These algorithms are usually light and easy to implement, which do not require the receiver to have expensive computing and data resources. The experimental results of different schemes are recorded in Table \ref{table:low} when we set the threshold as 32 dB.


\begin{table}[t]
    \centering
    \caption{The embedding rate of different schemes when the receiver owns non-learning-based image restoration algorithms. The threshold is 32 dB.}
    \label{table:low}
    \resizebox{0.95\linewidth}{!}{
    \begin{tabular}{cccccc}
    \toprule
    Method & BPDHE \cite{ibrahim2007brightness} & AMSR \cite{lee2013adaptive}  & LIME \cite{guo2016lime}  & NLFMT \cite{shreyamsha2013image}  & None  \\
    \hline
    DDH \cite{baluja2017hiding}         & 3$\times$24  & 3$\times$24  & 3$\times$24  & 3$\times$24  & 3$\times$24   \\
    UDH \cite{zhang2020udh}             & 2$\times$24  & 2$\times$24  & 2$\times$24  & 2$\times$24  & 2$\times$24   \\
    MISDNN \cite{das2021multi}          & 1$\times$24  & 1$\times$24  & 1$\times$24  & 1$\times$24  & 1$\times$24   \\
    HCVS \cite{weng2019high}     & 3$\times$24  & 3$\times$24  & 3$\times$24  & 3$\times$24  & 3$\times$24   \\
    LDH   & 16$\times$24 & 16$\times$24 & 16$\times$24 & 16$\times$24 & 16$\times$24  \\
    \bottomrule
    \end{tabular}
    }
\end{table}

\textbf{Receiver with Learning-based Algorithms.} A receiver may own expensive computing and data resources such that he can train a powerful image restoration model to enhance the image quality of the revealed secret images. In this experiment, we consider two state-of-the-art image restoration models: Restormer \cite{zamir2022restormer} and MPRNet \cite{zamir2021multi}, and train them respectively for different hiding schemes. We show the embedding rate of a high-capability receiver in Table \ref{table:middle-high}.

From both Table \ref{table:low} and \ref{table:middle-high}, we observe that the receiver with non-learning-based algorithms cannot increase the embedding rates of the involved schemes, while the receiver with learning-based algorithms can improve the quality of the secret images of the baselines, where Restomer is effective for both the lower and higher thresholds and MPRNet is effective for the lower threshold.
Our LDH has obvious advantages in terms of the embedding rates that are very high and are not affected by the receivers with different image restoration methods. The experimental results show that the involved baselines cannot hide more than 6 images when the threshold is larger than 26 dB, while our LDH can hide 64 images.

\begin{table}[t]
    \centering
    \caption{The embedding rate of different schemes when the receiver owns learning-based image restoration algorithms.}
    \label{table:middle-high}
    \resizebox{0.95\linewidth}{!}{
    \begin{tabular}{c|ccc|ccc}
        \toprule
        \multirow{2}{*}{Method} & \multicolumn{3}{c|}{PSNR$\geq$26dB} & \multicolumn{3}{c}{PSNR$\geq$32dB}  \\
        \cline{2-7}
                                & Restomer & MPRNet & None             & Restomer & MPRNet & None             \\
        \hline
        DDH \cite{baluja2017hiding}     & 6$\times$24     & 5$\times$24 & 5$\times$24               & 4$\times$24     & 3$\times$24 & 3$\times$24               \\
        UDH \cite{zhang2020udh}         & 5$\times$24     & 4$\times$24 & 3$\times$24               & 3$\times$24     & 2$\times$24 & 2$\times$24               \\
        MISDNN \cite{das2021multi}      & 4$\times$24     & 3$\times$24 & 2$\times$24               & 1$\times$24     & 1$\times$24 & 1$\times$24               \\
        HCVS \cite{weng2019high} & 6$\times$24     & 5$\times$24 & 5$\times$24               & 4$\times$24     & 3$\times$24 & 3$\times$24               \\
        LDH                     & 64$\times$24    & 64$\times$24 & 64$\times$24              & 16$\times$24    & 16$\times$24 & 16$\times$24              \\
        \bottomrule
    \end{tabular}
    }
\end{table}

\section{Conclusion}
\label{sec:conclusion}
In this paper, we propose a local deep hiding (LDH) scheme to demonstrate the reason behind the success of deep hiding at improving the embedding rate. LDH converts a secret image into a small imperceptible compact code using our hiding network and embeds the code into a random region of a cover image. The secret image can be revealed with a high quality using the locating and revealing network. Our extensive experiments have shown that the proposed method not only has a high embedding rate, but also achieves stronger robustness than existing deep hiding schemes. We also analyze the impact of different image restoration algorithms on the embedding rates of different deep hiding schemes.

\bibliographystyle{IEEEtran}
\bibliography{ref}

\end{document}